\documentclass[a4paper,11pt]{article}
\pdfoutput=1 

\usepackage{jinstpub} 

\makeatletter
\gdef\@fpheader{}  
\makeatother

\usepackage{lineno}

\title{\boldmath Learning Before Filtering: Real-Time Hardware Learning at the Detector Level}

\author[a]{Bo\v{s}tjan Ma\v{c}ek}

\affiliation[a]{Jo\v{z}ef Stefan Institute, Jamova cesta 39, 1000 Ljubljana, Slovenia}

\emailAdd{bostjan.macek@ijs.si}

\abstract{Advances in sensor technology and automation have ushered in an era of data abundance, where the ability to identify and extract relevant information in real time has become increasingly critical. Traditional filtering approaches, which depend on a priori knowledge, often struggle to adapt to dynamic or unanticipated data features. Machine learning offers a compelling alternative—particularly when training can occur directly at or near the detector.

This paper presents a digital hardware architecture designed for real-time neural network training, specifically optimized for high-throughput data ingestion. The design is described in an implementation-independent manner, with detailed analysis of each architectural component and their performance implications. Through system parameterization, the study explores trade-offs between processing speed, model complexity, and hardware resource utilization. Practical examples illustrate how these parameters affect applicability across various use cases.

A proof-of-concept implementation on an FPGA demonstrates in-situ training, confirming that computational accuracy is preserved relative to conventional software-based approaches. Moreover, resource estimates indicate that current-generation FPGAs can train networks of approximately 3,000 neurons per chip. The architecture is both scalable and adaptable, representing a significant advancement toward integrating learning directly within detector systems and enabling a new class of extreme-edge, real-time information processing.}

\newcommand{\imgScale}{0.28}

\begin{document}
\maketitle
\flushbottom

\section{Introduction}

For a long time, the majority of scientific fields were constrained by the inability to obtain a sufficient quantity of measurements. With the advent of automation and the rapid development of detectors, these limitations have largely disappeared. We now live in an era of data abundance, shifting the primary challenge to \emph{identifying the most relevant information} within an increasingly vast pool. This trend is expected to continue and intensify.

Traditionally, data filtering has relied on a-priori assumptions about the phenomena under investigation. However, the emergence of machine learning offers novel tools that can challenge and potentially reshape this paradigm. This research investigates the feasibility of embedding machine learning systems directly at or near the data sources. Such an approach would enable local extraction of relevant patterns from raw data, prior to any conventional filtering. A key advantage of this approach is the removal of biases introduced by preconceived assumptions, thereby allowing for the development of more effective \emph{data-driven filtering mechanisms} and consequently \emph{increasing discovery potential}.

Aiming for this goal, we are faced with a plethora of options, as machine learning offers a variety of distinct methodologies. However, if the final system is to be deployed at the detector level, a significant challenge arises: the absence of ground truth data. This limitation precludes the use of supervised learning and necessitates reliance on unsupervised approaches, narrowing the range of suitable methods. Among these, solutions based on neural networks stand out due to their flexibility and adaptability, making them strong initial candidates for investigation.

Several unsupervised learning paradigms based on neural networks are available, including generative adversarial networks, autoencoders, and normalizing flows — each capable of learning meaningful data representations without external labels. As the optimal choice among these depends strongly on the specific application, this study adopts a general perspective. It explores the feasibility of training a generic neural network architecture that can serve as a foundational building block for the aforementioned systems:
\begin{itemize}
\item Appropriately constrained, self-referential training enables the network to function as an autoencoder.
\item Parallel deployment of two networks supports the construction of generative adversarial networks.
\item Cascading smaller neural networks facilitates the implementation of normalizing flows.
\end{itemize}

The focus is thus on the \emph{design of training architecture} for fully connected neural network as the core computational element, rather than on the specifics of each individual unsupervised learning method. This approach enables the development of a broadly applicable hardware design that can later be specialized to meet the demands of particular application.

Training of neural networks is not a novel concept and is conventionally performed on Central Processing Unit (CPU) and Graphics Processing Unit (GPU) platforms. While these platforms are flexible and powerful, they are constrained by bandwidth, power consumption, and physical size—factors that significantly limit their integration into detector electronics, where the proposed system must ultimately reside. Consequently, this work presents an initial proposal for a digital hardware design designated for neural network training that is generic, adaptable, and suitable for a wide range of applications. The design is focused on maximizing the rate and volume of data that can be ingested during training, as this constitutes a primary bottleneck in pre-filter learning scenarios.

The goal of the presented study is to characterize the tradeoffs between data-intake speed, resource utilization, and the complexity of patterns that the system can extract. The design and results are discussed in an \emph{implementation-independent manner}, aiming to assess the overall feasibility of such an approach prior to any application-specific optimization. As such, the work should primarily function as a feasibility study, serving as a foundational step to guide future efforts in identifying applications where high-speed, on-device neural network training can be practically and effectively deployed.

\subsection{Motivation and Applications}

Numerous scientific domains stand to benefit from proposed design. Experiments in cosmology, gravitational wave detection, and High-Energy Physics (HEP) produce enormous volumes of data and exemplify the use of sophisticated filtering techniques. These fields represent ideal candidates for the integration of learning at the data source.

HEP experiments, in particular, offer a fertile ground for the deployment of such innovation. These experiments aim to deepen our understanding of fundamental particles and their interactions. Although the majority of known processes are well characterized, rare processes hold the potential to reveal new physics beyond the Standard Model. To maximize the likelihood of capturing these rare events, the rate of event production is increased. For instance, the current production rate at the Large Hadron Collider (LHC) yields an average of approximately $50$ overlapping processes every $25\,\mathrm{ns}$~\cite{LE2008}. Each such interval corresponds to a single event and produces a vast quantity of detector data~\cite{mlrates}.

To manage this data volume, experiments utilize sophisticated filtering mechanisms — referred to as triggers — that select events of interest while discarding the remainder. This process reduces the data rate to approximately $1000$ events per second~\cite{AtlasTrigger}, a manageable level for storage. As a result, a delicate balance must be maintained between production yeild and trigger precision.

Ad hoc theoretical assumptions about the nature of the universe are traditionally used to predict detector responses and build triggers that yield a high number of theoretically motivated events while reducing backgrounds. This approach has proven effective, but it also implies that our discovery potential is not just limited by experimental capabilities, but also by the human imagination in determining where to look.

The proposed method of learning before triggering holds significant potential in such context. By learning patterns from an \emph{unbiased sample of all generated events}, the system can extract meaningful structures directly from raw data. These learned representations can then be analyzed offline to identify regions of interest that may indicate novel phenomena. In this way, construction of the next generation trigger logic can be guided by data-driven insights rather than being solely constrained by theoretical assumptions—thus expanding the discovery potential beyond the limits of current paradigms.

\subsection{Focus of Research and Related Work}

Machine learning (ML) has become a pervasive technological paradigm, with widespread applications across diverse fields—including HEP. The work presented in this paper positions itself within this domain by focusing on \emph{hardware acceleration} for ML, specifically targeting the \emph{training stage} of models in data-rich and \emph{resource-constrained environments} such as particle detectors.

For a general overview of ML accelerators and their evolution please refer to the comprehensive review~\cite{GUPTA20211}, which outlines the trajectory from simple computational speedups toward more application-specific requirements like low-latency and low-power processing, particularly for on-edge ML deployments. In edge environments like detectors, constraints on power, form factor, and cooling capacity necessitate highly efficient architectures.

Modern acceleration platforms typically fall into four categories: GPUs, Tensor Processing Units (TPU), Field Programmable Gate Arrays (FPGA), and Application Specifc Integrated Circuits (ASIC). A recent comparative analysis of these platforms is provided in~\cite{Burhanuddin_2023}, which highlights the dominance of GPUs in contemporary acceleration pipelines due to their flexibility, widespread software support, and suitability for highly parallel tasks. FPGAs, while less general-purpose, are highly reconfigurable and significantly more energy-efficient under certain conditions, making them ideal for prototyping and domain-specific adaptation. This energy efficiency in the training context is emphasized by~\cite{Tao_2020_CVPR_Workshops}, particularly in the context of limited-resource platforms. Once an accelerator design matures and warrants deployment at scale, ASICs provide unmatched performance and customization flexibility. However, their lack of reconfigurability makes them suitable mainly for stable environments. Insightful overviews of ASIC implementations for inference are available in~\cite{MOOLCHANDANI2021101887,MACHUPALLI2022104441}, while~\cite{CHEN2020264} offers a broader perspective across multiple accelerator paradigms.

A large-scale review of FPGA-based acceleration research can be found in~\cite{yan2024surveyfpgabasedacceleratorml} that analyzed 287 articles and provides valuable empirical insight into the state of the field. It shows a strong imbalance: 81\% of research is centered on inference acceleration, while only 13\% is concerned with accelerating training. This discrepancy highlights a key research gap: hardware-accelerated training remains underexplored and immature, particularly in closed or edge environments where flexibility and autonomy are critical.

\subsubsection*{Contribution of This Work}

The work presented here directly addresses this gap. It diverges from mainstream developments by aligning itself with respect to two major challenges:

\begin{itemize}
  \item \textbf{Efficient training acceleration:} This field remains significantly understudied. While~\cite{Burhanuddin_2023} touches on training, it is predominantly concerned with architectural differentiation across platforms. A more dedicated discussion on training accelerators can be found in~\cite{10940371}, with an earlier but still relevant synthesis in~\cite{fi12070113}. This paper presents work focused specifically on the training phase as its primary objective.

  \item \textbf{Autonomy of training systems:} Current efforts in hardware acceleration for training predominantly target individual components such as matrix multiplication, backpropagation, and weight updates rather than full-stack training systems. Work presented in~\cite{10.1145/3174243.3174258} presents a configurable matrix multiplication framework for deep learning workloads on FPGAs, while~\cite{10.1145/3307650.3322237} proposes FloatPIM, an in-memory training accelerator designed to reduce data movement and enhance speed and energy efficiency. From an algorithmic standpoint~\cite{JMLR:v24:22-1208} provides a compelling discussion of the algorithmic speedup problem, advocating for bottleneck-aware architectural co-designs. While each of these approaches has demonstrated significant improvements, the presented work emphasizes a system-wide optimization of all training components to produce an \emph{autonomous training primitive} that can be integrated into larger designs.
\end{itemize}

Together, the cited works underscore a common theme: acceleration efforts prioritize global parallelism or optimized primitives within broader heterogeneous systems. By contrast, this research proposes a self-contained architecture—a co-designed blend of compute and memory logic—that supports autonomous training close to the data source. This enables early and efficient selection of relevant features or events before data is further processed or transmitted. As this is the main focus of the presented work, detailed application-specific optimization will not be discussed. The results aim to capture trade-offs between system scale, resource usage, and data absorption speed in order to inform the future research on feasibility of different concrete applications.

\subsubsection*{Relevance to HEP Applications}

While the conclusions of this research are broadly applicable, they offer particularly valuable insights for HEP. A broader review of fast ML in scientific contexts, including HEP, is provided in~\cite{10.3389/fdata.2022.787421}, where the relevance and transformative potential of embedded ML systems are clearly visible.

Today, ML is deeply embedded in the analysis pipelines and computational infrastructure of major HEP experiments~\cite{hepMLinfrastructure}. Concrete examples include ML applications in neutrino experiments~\cite{QIAN2021165527}, where event classification and signal extraction tasks benefit from advanced inference models. More recently, ML has begun to permeate detector design itself~\cite{duarte2022fastmlsciencebenchmarksaccelerating}; however implementations at the detector level, thus far have remained limited to inference-only models.

Among the most urgent application of fast ML in HEP hardware is sensor data compression (content-loss reduction). It is motivated by the enormous data throughput generated by modern experiments—often exceeding 100~TB/s~\cite{bartoldus2022innovationstriggerdataacquisition}. With data transmission bandwidth representing a fundamental bottleneck, several parallel strategies are being pursued. One is the development of on-detector intelligence for real-time data compression, e.g.~\cite{201925,8424231,AXIOTIS}. Another strategy is object reconstruction, where raw detector signals are transformed into high-level physics objects. Tasks such as particle tracking and identification~\cite{doi:10.1142/9789811234033_0012}, jet classification~\cite{bileska2025designfpgaimplementationwombat}, vertex reconstruction, and calorimetric measurements~\cite{inferenceFPGAhep} are being pushed closer to the front-end layer. In each case, the goal is to extract and transmit only the most relevant physics information, thereby minimizing data volumes at the earliest possible stage.

A third and increasingly active strategy is \emph{data reduction through intelligent filtering} (content-selection reduction). Early successes include the CICADA system~\cite{CMS-DP-2024-121} and the 40~MHz autoencoder-based trigger~\cite{autoencoder40LHC}, both of which demonstrate the feasibility of ML-driven decision-making at the edge. Nevertheless, current systems are still based on pre-trained inference models.

In contrast to the approaches described above, which rely on offline learning to support detector inference, this study explores a novel direction: integrating \emph{learning within the detector} itself. By investigating the characteristics and limitations of the fundamental building block required for this task, this work supports further development of real-time dynamic pattern recognition and adaptive feature selection. While the ability to learn complete physics processes remains beyond the capabilities of current hardware, this study aims to develop and characterize an \emph{autonomous training primitive} with high data absorption, suitable for integration at any layer of the detector.

Its high data absorption is a critical asset in the front-end electronics of high-throughput environments such as the LHC~\cite{LE2008,AtlasTrigger}, where early information extraction at the level of segmented detectors can substantially reduce downstream data complexity and storage demands. At the same time, the autonomous nature of the proposed design enables its application in higher-level detector services, such as the trigger system, where data collected from all subsystems must be processed efficiently.

\subsection{Paper Organization}

As motivated, this work focuses on the training of a fully connected neural network as a baseline. Such networks are known for their versatility and serve as a fundamental, general-purpose building block in many AI systems. 

Section~\ref{sec:prepare} introduces a minimal set of equations that govern the training process and outlines the learning procedure. Sections~\ref{sec:overview} and~\ref{sec:principles} describe the high-level architecture and operational principles of the proposed design. Section~\ref{sec:components} details the individual design components, providing sufficient information to parameterize the system and identify the constraints they impose.

With these constraints established, Section~\ref{sec:absorbtion} presents a methodological evaluation of system performance in terms of data absorption capability. This section includes \emph{the core results of the study—an applicability landscape}, with respect to model complexity, system scale, processing speed, and resource limitations.

Section~\ref{sec:verification} presents a proof-of-principle demonstration, validating the proposed design through practical implementation. This is followed by discussion in Section~\ref{sec:dicsussion}, which places the results in the context of current challenges and articulates their broader impact. Finally, Section~\ref{sec:implementation} outlines future research directions that naturally emerge from this work.

\subsection{Nomenclature and Definitions}

To aid reading about this rather complex topic, this short section introduces a few conventions that are followed throughout the paper to simplify understanding. The system being described is parameterized and designed to learn from data. In this notation:

\begin{itemize}
    \item Values of the actual data being processed by the system, as well as transformation functions, are denoted using Greek letters (e.g., $\alpha$, $\Sigma$).
    \item Indexes and system parameters are represented by Latin letters. Lowercase letters (e.g., $n$) denote parameters that can be reconfigured during system operation - \emph{run-time parameters}. In contrast, parameters that must be selected during system design are denoted by uppercase letters (e.g., $N$) - \emph{construction-time parameters}. When a derived value is represented by a lowercase letter, it indicates that at least one of the parameters influencing it is a run-time parameter, otherwise an uppercase letter is used.
\end{itemize}

\begin{figure}[htbp]
    \centering
    \includegraphics[scale=\imgScale]{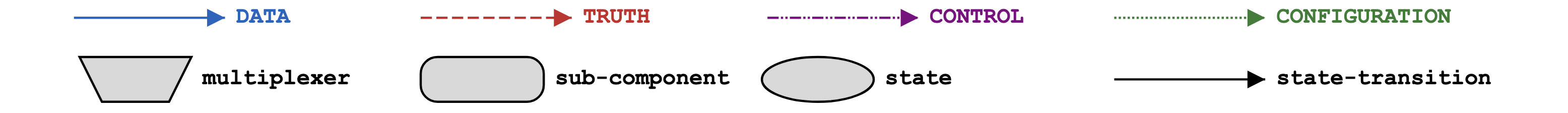}
    \caption{Common legend for upcoming figures throughout the paper.}
    \label{fig:legend}
\end{figure}

This convention will help in keeping track of the constraint type being referred to at any point along the text. Additionally, the paper includes figures depicting state transitions and component substructures. To avoid redundancy, a common legend is provided in Figure~\ref{fig:legend}, which explains the shapes, colors, and line styles used throughout the figures.

\subsection{Preparing Neural Networks for Learning in Hardware}
\label{sec:prepare}

\begin{figure}[htbp]
    \centering
    \includegraphics[scale=\imgScale]{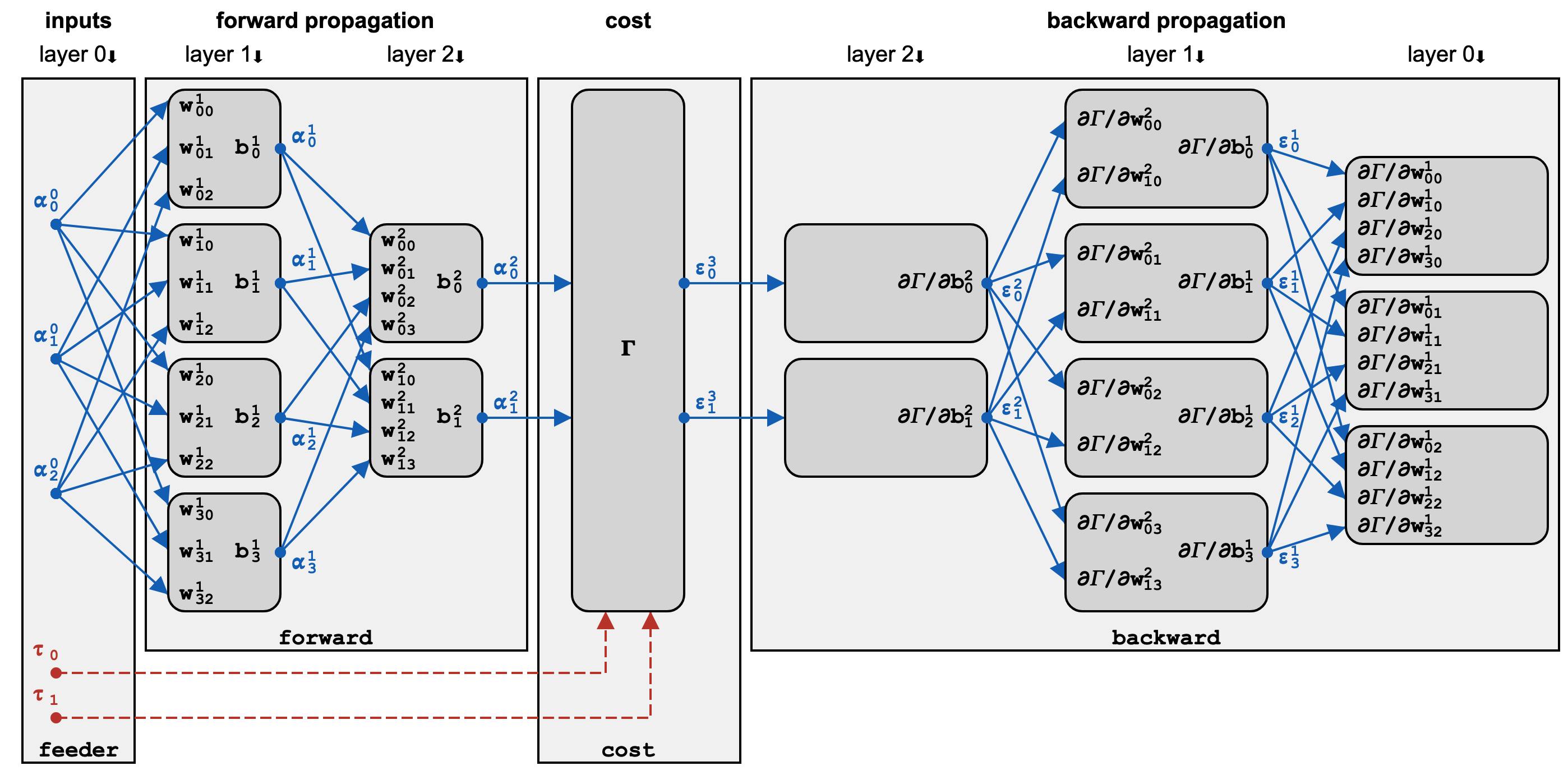}
    \caption{Computational flow of a two-layer network with 4 neurons in the first and 2 neurons in the second layer. The process flows from left to right. Inputs $\alpha_j^0$ are forward propagated, yielding the network's prediction $\alpha_j^2$. Each box in this section represents the computations required by one neuron and includes the network parameters necessary for that computation. The cost function $\Gamma$ then evaluates the deviations from the expected results $\tau_j$. These deviations feed backward propagation. In this phase, each box corresponds to set of computations, now populated with the calculated gradients of the cost function. All computational steps are distributed among the \texttt{feeder}, \texttt{forward}, \texttt{cost}, and \texttt{backward} blocks, which correspond to the design’s subcomponents. These are listed here for completeness and will be referenced again in later sections.}
    \label{fig:math}
\end{figure}

The details and intricacies of neural networks and their training are extensively covered in~\cite{DeepLearning}. However, to understand the key aspects of the proposed design, a simplified mathematical explanation suffices. The core premise is that training consists of three main steps: making a prediction based on input data (typically referred to as forward propagation), evaluating this prediction (comparison to desired result) and estimating the adjustments needed to improve it (referred to as backward propagation). The full computational flow required to train the network is visualized in Figure~\ref{fig:math}.

The mathematics of forward propagation is based on the fact that neurons are arranged in layers, with each neuron's activation (i.e., its output) in one layer serving as input to the neurons in the subsequent layer. In a fully connected topology, each neuron in layer $l$ is influenced by all activations from the previous layer $l-1$:
\begin{equation}
    \chi_j^l = \sum_k (w_{jk}^l \cdot \alpha_k^{l-1}) + b_j^l.
\label{eq:fwdn_stimulus}
\end{equation}
Here, superscript indices denote layer numbers, and subscript $j$ refers to a specific neuron in layer $l$. The index $k$ enumerates neurons in layer $l-1$, and their activations $\alpha_k^{l-1}$ are summed, satisfying the fully connected condition. Each term is weighted by $w_{jk}^l$, and the sum is offset by the bias $b_j^l$. These weights and biases are collectively referred to as the network’s parameters. The result $\chi_j^l$, called the neuron's stimulus, determines its output activation:
\begin{equation}
    \alpha_j^l = \sigma^l(\chi_j^l),
\label{eq:fwdn_activation}
\end{equation}
where $\sigma^l$ denotes the activation function used in layer $l$. These activations are then passed to the next layer. The entire process starts by absorbing an external data \emph{sample, i.e. a collection of input values}. These can be treated as the activations of layer $l=0$, and are thus denoted as $\alpha_j^0$. Using equations~\eqref{eq:fwdn_stimulus} and~\eqref{eq:fwdn_activation}, the network’s response can be computed layer by layer until reaching the final layer $L$, where $\alpha_j^L$ represents the network’s output\footnote{The same computations are used when the pre-trained network is deployed for inference.}.

Training the network involves adjusting $w_{jk}^l$ and $b_j^l$ to improve the network's predictions. This is typically done using gradient descent: the network’s final prediction is compared to the desired result $\tau_j$ (referred to as the “truth”), and the difference is quantified by a cost function $\Gamma$. The stimulus error in the final layer is based on derivative of the cost function with respect to the neuron's activation:
\begin{equation}
    \epsilon_j^{L} = \left(\frac{d\Gamma(\alpha^L_j, \tau_j)}{d\alpha_j^L}\right) \cdot \sigma'^{L}(\chi_j^L).
\label{eq:bckn_propagation_last}
\end{equation}
This error can then be propagated backward through the layers:
\begin{equation}
    \epsilon_j^l = \sum_k (w_{kj}^{l+1} \cdot \epsilon_k^{l+1}) \cdot \sigma'(\chi_j^l).
\label{eq:bckn_propagation_others}
\end{equation}
In both equations, $\sigma'$ is the derivative of the activation function, and $\frac{d\Gamma}{d\alpha_j^L}$ is the derivative of the cost function with respect to the prediction\footnote{This is $\epsilon_j^{L+1}$, which equals $\epsilon_j^{3}$ in the context of Figure~\ref{fig:math}.}.

From the computed error terms, the gradients of the cost function with respect to the network parameters are:
\begin{equation}
    \frac{\partial\Gamma}{\partial b_j^l} = \epsilon_j^l,
\label{eq:bckn_gradb}
\end{equation}
\begin{equation}
    \frac{\partial \Gamma}{\partial w_{jk}^l} = \alpha_k^{l-1} \cdot \epsilon_j^l.
\label{eq:bckn_gradw}
\end{equation}
These gradients estimate how much each weight and bias should be adjusted to reduce the cost:
\begin{equation}
    b_j^l \rightarrow b_j^l - s \cdot \frac{\partial\Gamma}{\partial b_j^l},
\label{eq:update_stepb}
\end{equation}
\begin{equation}
    w_{jk}^l \rightarrow w_{jk}^l - s \cdot \frac{\partial\Gamma}{\partial w_{jk}^l},
\label{eq:update_stepw}
\end{equation}
where $s$ is the learning rate — a parameter that controls the step size during gradient descent. If appropriately chosen, repeated application of this update rule leads the parameters to converge to optimal values.

The equations above define the complete set of computations required to train a neural network. They will serve as the mathematical justification for the structure of the proposed design in the sections that follow. Note that both forward and backward propagation rely heavily on Multiply-And-Accumulate (MAA) operations, with backward propagation being computationally more intensive due to the extra steps required for gradient calculation~\eqref{eq:bckn_gradw} and parameter updates~\eqref{eq:update_stepb}, \eqref{eq:update_stepw}.

\section{Design overview}
\label{sec:overview}

\begin{figure}[htbp] 
    \centering 
    \includegraphics[scale=\imgScale]{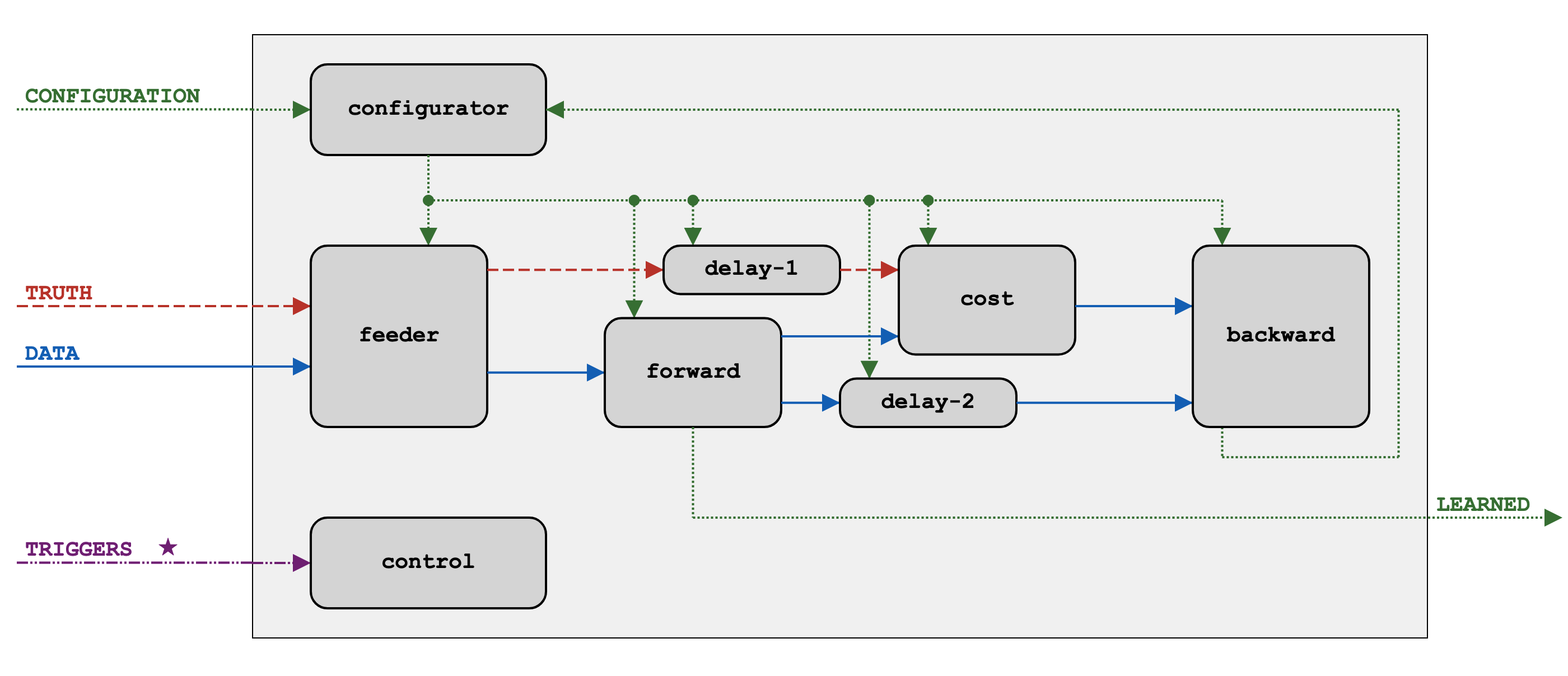} 
    \caption{Structural design of the proposed system. Signals marked with $\bigstar$ influence internal state transitions. Note the pipeline structure in the data flow and the resulting ability of components to compute in parallel.} 
    \label{fig:design} 
\end{figure}

As discussed in the previous section, training a neural network using linear gradient descent involves the following steps for each input sample: forward propagation, cost calculation, backward propagation, and gradient computation. To maximize input bandwidth, these computations should be performed in parallel. Accordingly, each of the aforementioned steps is handled by a dedicated component, all operating concurrently. Figure~\ref{fig:design} illustrates the overall system architecture. Data moves through the system in a pipelined fashion: while the next sample is being prepared by the \texttt{feeder}, the \texttt{forward} block processes the current sample. Simultaneously, the \texttt{cost} and \texttt{backward} blocks handle increasingly older samples to perform gradient evaluation. Because multiple samples are processed concurrently at different stages of the pipeline, delay elements (\texttt{delay-?}) are introduced to ensure data alignment when the sample information is later merged in a common processing unit.

In addition to the main learning path, two auxiliary service components are included. The first, \texttt{configurator}, handles the initial configuration and updates learning parameters as needed. The second, \texttt{control}, manages the internal states of the various components, ensuring coordinated operation across the system.

This architecture is driven by the goal of keeping all submodules active during data intake, thereby optimizing the sample processing rate. Consequently, the latency introduced by pipeline traversal is of secondary importance.

\section{Principles of operation}
\label{sec:principles}

\begin{figure}[htbp]
    \centering 
    \includegraphics[scale=\imgScale]{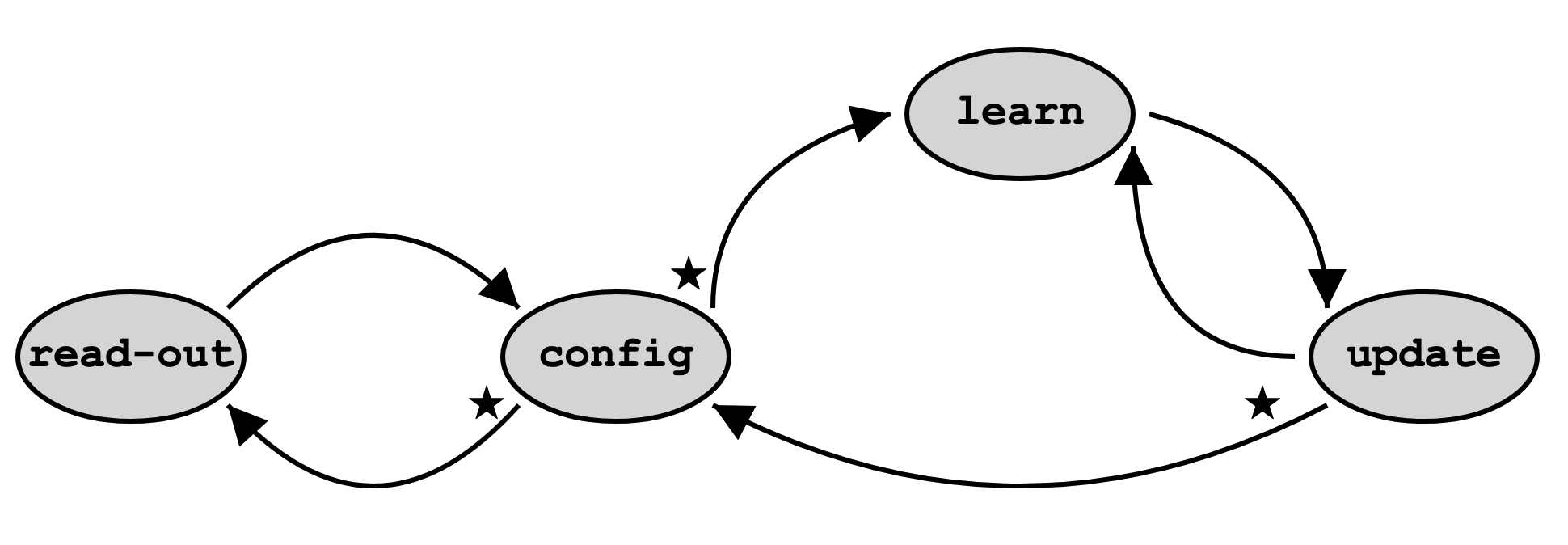} 
    \caption{Design's operational states and transitions. Transitions marked with $\bigstar$ are user triggered, those without are automatic transitions based on internal conditions.} 
    \label{fig:states} 
\end{figure}

The operational states of the design are illustrated in Figure~\ref{fig:states}. After a reset, the desing initializes in the \texttt{config} state. In this state, it is sensitive to the \texttt{CONFIGURATION} input, which is responsible for setting network-wide parameters, as well as configuring both the \texttt{forward}, \texttt{backward} and \texttt{cost} blocks (see Figure~\ref{fig:design}).

Once the network is fully configured, it responds to the 'start-learning' trigger, transitioning to the \texttt{learn} state. During this phase it receives data and corresponding truth values via the \texttt{DATA} and \texttt{TRUTH} inputs. The \texttt{feeder} module accumulates these inputs until a complete sample is formed, after which the data is streamed into the \texttt{forward} block, which in turn generates the network’s prediction. The \texttt{cost} module then compares this prediction to the truth, computing the cost derivatives for the given sample. These derivatives are subsequently used by the \texttt{backward} block to perform backward propagation and compute gradients.

It is standard practice to repeat this procedure for a small collection of samples, referred to as a batch, and compute the average of the calculated gradients. The number of samples within a batch is hereafter denoted by $n_{\mathrm{batch}}$. Once the entire batch has been processed, the network transitions to the \texttt{update} state. In this state, the \texttt{backward} block applies the accumulated gradients to update the network parameters, thereby reconfiguring the entire design and preparing it for the next batch.

Once the final reconfiguration command has propagated through the last corner of the design, the system checks for the presence of the 'end-of-learning' trigger. If this trigger has been issued, the network transitions back to the \texttt{config} state, where it remains idle until further instructions are received. Otherwise, the learning process continues for another batch in \texttt{learning} state.

When the design is in the \texttt{config} state, the user can initiate a readout of the trained weights and biases by issuing 'start-of-readout' trigger. In response the system transitions to \texttt{read-out} state and outputs a configuration stream via the \texttt{LEARNED} output, providing the current parameter values. Once readout is complete design returns to the \texttt{config} state, from where it can either be reconfigured or resume the learning process.

\section{Submodule Design and Functionality}
\label{sec:components}

To maximize design's utility, it must be highly configurable. To achieve an optimal balance between flexibility and performance, only hyper-parameters whose reconfiguration would severely limit system performance are predefined. These hyper-parameters include the number of layers $N_{\mathrm{layers}}$ and the maximum number of neurons within each layer $N_{\mathrm{neurons}}$. With these two construction-time hyper-parameters, a topological grid is established, providing constraints within which any neural network can be configured at runtime.

Complementing these are hyper-parameters configurable at runtime. The first is the network shape, defined by the set of numbers of active neurons\footnote{Those not selected to be active are functionally transparent, acting merely as simple data pass-through modules. As such, they will not be discussed further.} $n^l \leq N_{\mathrm{neurons}}$ within each layer $l$. In addition the user must also define the batch size $n_{\mathrm{batch}}$ and the step size $s$ used in gradient descent.

With all of these chosen, the structure and operation of the system is detailed in the following subsections. The primary goal is to outline the system's limitations, which later serve as the basis for characterizing the design. It is important to note again that, while the design is presented here in an \textit{implementation-independent} manner, it is assumed to be a digital logic design. Consequently, its speed performance metrics should scale with the frequency of the underlying digital clock, up to the limit imposed by the slowest component. To that end, all time-related discussions that follow are expressed in units of the underlying clock period.

\subsection{Data feeder}

To decouple the external environment from the design's internal operation as much as possible, the \texttt{feeder} performs two key tasks.

The first task is to ensure that the entire sample is available before processing it. All samples must be of uniform size, containing $n_{\mathrm{inputs}}$ inputs, with each input represented by a single numerical value. Data is passed to the \texttt{forward} block only when a complete sample is assembled and the network is in operational mode.

The second task is rate averaging. Since the design targets data sources with a continuous flow, a challenge arises during the \texttt{update} state (see Figure~\ref{fig:states}), when the network pauses input consumption to update its parameters. To accommodate this, the \texttt{feeder} incorporates a small memory buffer implemented as a First-In-First-Out (FIFO) structure. This buffer accumulates data during the \texttt{update} state and allows for faster consumption during the subsequent \texttt{learn} state. It is critical that the FIFO be large enough to cushion the absorption fluctuations caused by these learning cycles. The required FIFO depth depends on the duration of the update cycles and the mismatch between input and consumption bandwidths.

\subsection{Forward network}

It is clear from the iterative nature of Equation~\eqref{eq:fwdn_stimulus} that the task of forward propagation can be done for all layers concurrently. Thus, while the forward network is part of a larger pipeline, it can itself be structured as a pipeline of forward layers. This subsection first discusses two basic building blocks: the forward neuron and the activation structure. Understanding these two subcomponents enables us to discuss the forward layers and the limits they impose on the overall design.

\subsubsection{Forward neuron}

\begin{figure}[htbp] 
    \centering 
    \includegraphics[scale=\imgScale]{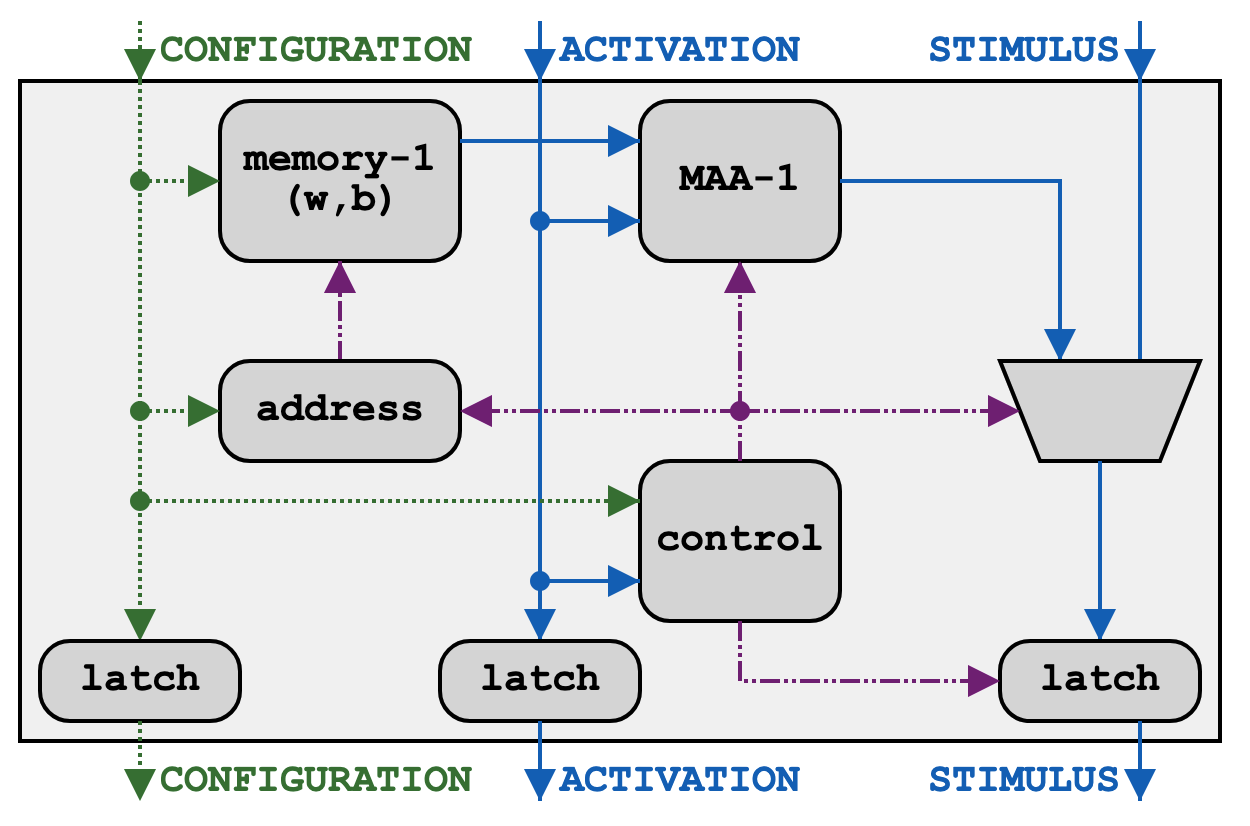} 
    \caption{Internal structure of forward neurons. The main computation is performed by the \texttt{MAA-1} unit using the local \texttt{memory-1} to appropriately weigh the activations from the previous layer and perform biasing. The stimulus, which is the unit's final result is presented at \texttt{STIMULUS} output.} 
    \label{fig:fwdneuron} 
\end{figure}

The forward neuron is the basic building block tasked with calculating the neuron's stimulus. Its internal structure is depicted in Figure~\ref{fig:fwdneuron}. Its central component, \texttt{MAA-1}, performs one MAA operation of Equation~\eqref{eq:fwdn_stimulus} on each clock cycle. The activations $\alpha_k^{l-1}$ are supplied by the neuron's \texttt{ACTIVATION} input stream, one per clock cycle. Network parameters are supplied by the local memory unit \texttt{memory-1}. During configuration, the \texttt{CONFIGURATION} stream pre-fills this memory at consecutive addresses with weights, immediately followed by the bias at the last address used. Since one MAA operation is executed on each clock cycle, the addition of the bias is achieved by injecting its value along with a fixed value of 1 in the \texttt{ACTIVATION} stream. Thus, an extra fixed activation of 1 must be appended to the activation stream that is passed from the previous layer. Once this is absorbed, the \texttt{MAA-1} introduces a latency of $D_{\mathrm{fwd}}$ before the final accumulation is output on its output port. As this latency is deterministic, the \texttt{control} structure appropriately latches the final stimulus to the neuron's \texttt{STIMULUS} port.

It should also be highlighted that the \texttt{control} structure requires an additional clock cycle to reset the neuron's sub-components. Combined with the fixed trailing activation, this imposes a limit of two clock cycles between subsequent samples being processed.

Lastly, it is important to note that each neuron can be configured in transparent mode, effectively disabling all of its functionality except for pipelining the \texttt{CONFIGURATION} and \texttt{ACTIVATION} streams. This feature allows for the selective use of a subset of neurons, thereby defining the shape of the neural network dynamically at runtime.

\subsubsection{Activation module}

\begin{figure}[htbp] 
    \centering 
    \includegraphics[scale=\imgScale]{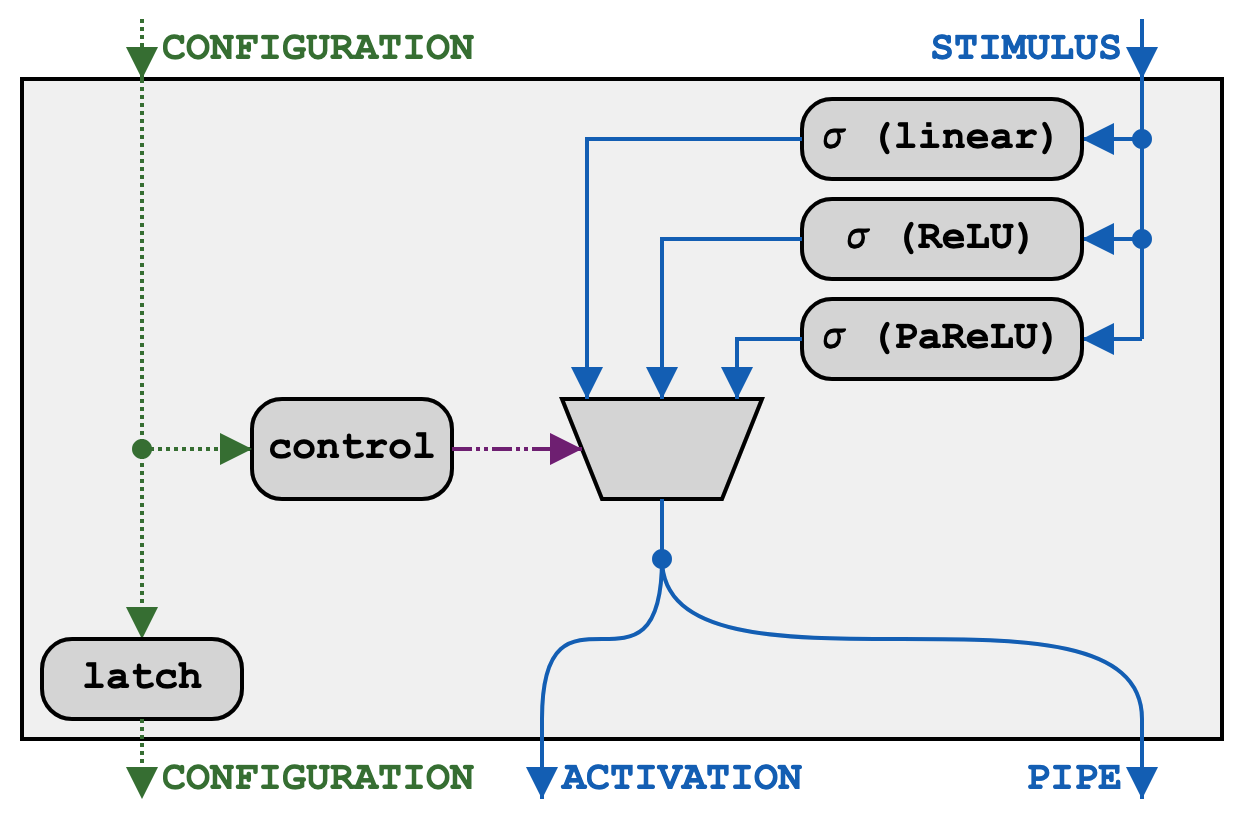} 
    \caption{Internal structure of the activation module whose task is to process the stimuli produced by the forward neurons of the same layer and compute their activations. It provides runtime configurable selection from a predefined list of implemented activation functions. As a byproduct, it also computes activation derivatives, which are served by the \texttt{PIPE} output and later used in backward propagation.} 
    \label{fig:fwdactivation} 
\end{figure}

As forward neurons consume the \texttt{ACTIVATION} stream, they produce the \texttt{STIMULUS} stream. These stimuli correspond to neurons in a particular layer and are processed by the activation module. It provides dual functionality: it calculates the activations \eqref{eq:fwdn_activation} and the activation derivatives $\sigma'^l(\chi_j^l)$, which are later used in backward propagation (within Equation~\eqref{eq:bckn_propagation_others}). The functional schematic is depicted in Figure~\ref{fig:fwdactivation}, which illustrates multiple activation blocks instantiated in parallel. Such design allows for individually configurable activations for each layer, broadening its usability. The same \texttt{CONFIGURATION} stream, used with forward neurons, can also be used to select one of the computed activations and their derivatives. The \texttt{ACTIVATION} output provides only activations, while the \texttt{PIPE} output holds both activations and their derivatives.

To allow for high-speed processing, activation blocks must be implemented with equal and fixed computational latency $D_{\sigma}$. The three functions depicted—linear, ReLU~\cite{relu}, and PaReLU~\cite{relu}—serve as functionally sufficient examples of activations that can be efficiently computed in a pipelined manner within three clock cycles.

\subsubsection{Forward layer}

\begin{figure}[htbp] 
    \centering 
    \includegraphics[scale=\imgScale]{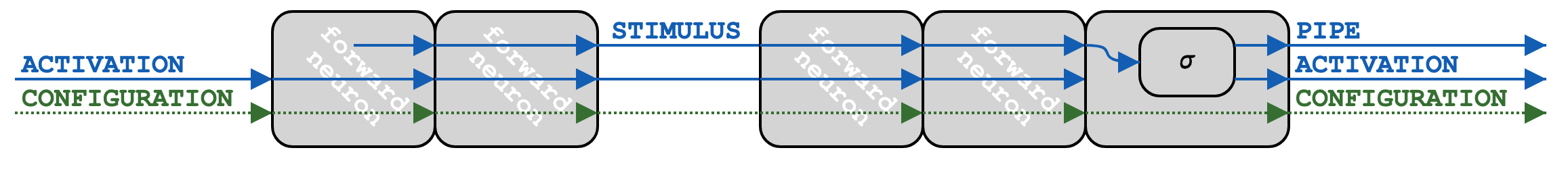} 
    \caption{The internal structure of the forward layer. A pipeline of forward neurons is followed by the activation module, which transforms neuron stimuli into activations. These activations serve as inputs to an identical structure of the subsequent layer. Data progresses through the layer at a rate of one clock cycle per neuron, effectively enabling one MMA operation at each neuron in every clock cycle.} 
    \label{fig:fwdlayer} 
\end{figure}

With the components defined, the forward layer is constructed as a pipeline of forward neurons followed by an activation module, as depicted in Figure~\ref{fig:fwdlayer}. Both the \texttt{CONFIGURATION} and \texttt{ACTIVATION} streams propagate from one neuron to the next on every clock cycle. As input activations arrive, neurons perform their computations and calculate their stimuli. These values are latched into the \texttt{STIMULUS} pipeline. Only once results of all active neurons within the layer are available, can the data be pushed along the pipeline and consumed by the activation structure. This structure produces activation values, which serve as inputs to the next layer. Simultaneously, the \texttt{PIPE} stream (holding both activations and derivatives) is generated and eventually fed into the \texttt{delay-2} (see Figure~\ref{fig:design}).

A key advantage of this neuron arrangement is its efficiency: one MAA operation is performed within each neuron on every clock cycle during the traversal of a sample. However, there are two implications of such a structure:
\medskip

\textbf{Implication 1:} Lets start by setting the time reference to just before the first input of the first sample enters the layer. Initially, it takes $N_{\mathrm{neurons}} - n^l$ time steps for the first input to reach the first active\footnote{The design assumes that all transparent neurons are clustered at the start of the layer chain.} neuron. If there are total of $n_{\mathrm{inputs}}^l$ inputs per sample, the first active neuron requires $N_{\mathrm{neurons}} - n_l + n_{\mathrm{inputs}}^l + D_{\mathrm{fwd}}$ steps to produce and latch its stimulus onto the \texttt{STIMULUS} stream. Since each subsequent neuron lags by one clock cycle, the last neuron in the layer produces its result at step $N_{\mathrm{neurons}} + n_{\mathrm{inputs}}^l + D_{\mathrm{fwd}}$. Only at this point are all stimuli of the layer ready, thus allowing propagation along the \texttt{STIMULUS} stream to begin—this includes the result of the first active neuron in the layer.

Now, assume that the second sample starts after $p_{\mathrm{sample}}^l$ clock cycles with respect to the start of the first sample. The first active neuron will produce its second stimulus at $p_{\mathrm{sample}}^l + N_{\mathrm{neurons}} - n_l + n_{\mathrm{inputs}}^l + D_{\mathrm{fwd}}$.

To prevent data loss, this must occur after the stimulus of the first sample has already been shifted away. This condition imposes the following constraint on period with which layer can digest samples:

\begin{equation}
    p_{\mathrm{sample}}^l \geq \max(n_{\mathrm{inputs}}^l + 2, n_l),
\label{eq:pause_layer}
\end{equation}

where the lower bound of $n_{\mathrm{inputs}}^l + 2$ is dictated by the requirement that the forward neuron must have at least two clock cycles of rest between subsequent samples, as described in previous subsections. Since the processing speed must adapt to the slowest layer, we must determine the maximum value of $p^l_{\mathrm{sample}}$ across all layers. Given that in a fully connected neural network $n_{\mathrm{inputs}}^l = n^{l-1}$ and assuming that at least one layer is fully utilized\footnote{If this assumption does not hold, an alternative design with an appropriately smaller $N_{\mathrm{neurons}}$ can be considered, and performance can be evaluated accordingly.}, this yields the overall period constraint, that samples entering the desing must obey:

\begin{equation}
    p_{\mathrm{sample}}^{\mathrm{tot}} \geq \max(n_{\mathrm{inputs}}, N_{\mathrm{neurons}}) + 2.
\label{eq:pause}
\end{equation}

This result implies that when the network contains more neurons in the largest layer than the number of inputs per sample, additional dead time between input samples must be accounted for—representing one of the constraints on input bandwidth.

\textbf{Implication 2:} The second variable to analyze is data propagation latency. In the previous paragraph, we noted that $N_{\mathrm{neurons}} + n_{\mathrm{inputs}}^l + D_{\mathrm{fwd}}$ clock cycles are required for stimulus computation of the last neuron in the layer. Thus, the total latency induced by a forward layer is:
\begin{equation}
    l_{\mathrm{fwd}}^l = N_{\mathrm{neurons}} + n^{l-1} + D_{\mathrm{fwd}} + D_{\sigma}
\label{eq:latency_fwd}
\end{equation}

Both of these implications will later be used to estimate the overall design performance limitations.

\subsection{Cost structure}
\label{sec:cost}

The previous subsection detailed the functionality of the forward network, which produces the network’s prediction. The role of the cost module is to evaluate the deviation of this prediction from the desired outcome, in the form of the cost derivative $\frac{d\Gamma}{d\alpha_j^L}$. An approach similar to that used in the activation structure can be employed here: multiple cost functions can be implemented and computed concurrently. However, it is crucial that all cost functions exhibit the same deterministic latency, denoted $D_{\mathrm{cost}}$. If this condition is satisfied, the \texttt{CONFIGURATION} stream can be used to select the desired cost function at runtime.

For the general conclusions presented in the following chapters, the choice of cost function does not affect the results. The cost function should be chosen according to the purpose of the application, while ensuring that the latency requirement $D_{\mathrm{cost}}$ is met. One of the simplest options is the sum-of-squares, $\Gamma = \sum_j (\alpha_j^L - \tau_j)^2$, which exhibits two useful properties. First, its derivative is a simple subtraction, which can be implemented as a low-latency operation. Second, information between different output neurons is not mixed, thus simplifying the design.

\subsection{Delay lines}

Equation~\eqref{eq:bckn_propagation_others} states that backward propagation requires activations and their derivatives, which are computed in the forward network. As described, it is the activation structure that generates both values and streams them via the \texttt{PIPE} output.

Once these results are produced, the underlying sample information propagates further through the rest of the forward network, the cost function, and into the respective part of the backward network. Since all of these components are implemented as pipelines, the activations and derivatives must be delayed by an equal amount to ensure they reach the backward propagation component in sync with the sample information. This synchronization is achieved using delay lines. Their effective delay is configurable via the \texttt{CONFIGURATION} stream and must be set at runtime to maintain the required synchronization for proper design operation.

\subsection{Backward network}

Examining Equations~\eqref{eq:fwdn_stimulus} and~\eqref{eq:bckn_propagation_others}, it becomes evident that, with the additional multiplication factor, backward propagation can be performed in a manner similar to forward propagation. The same reasoning that justified the division of forward propagation into sub-pipelines applies equally to backward propagation. To that end, we first discuss the structure of backward neurons and the update module before discussing the layers themselves.

\subsubsection{Backward neuron}

\begin{figure}[htbp] 
    \centering 
    \includegraphics[scale=\imgScale]{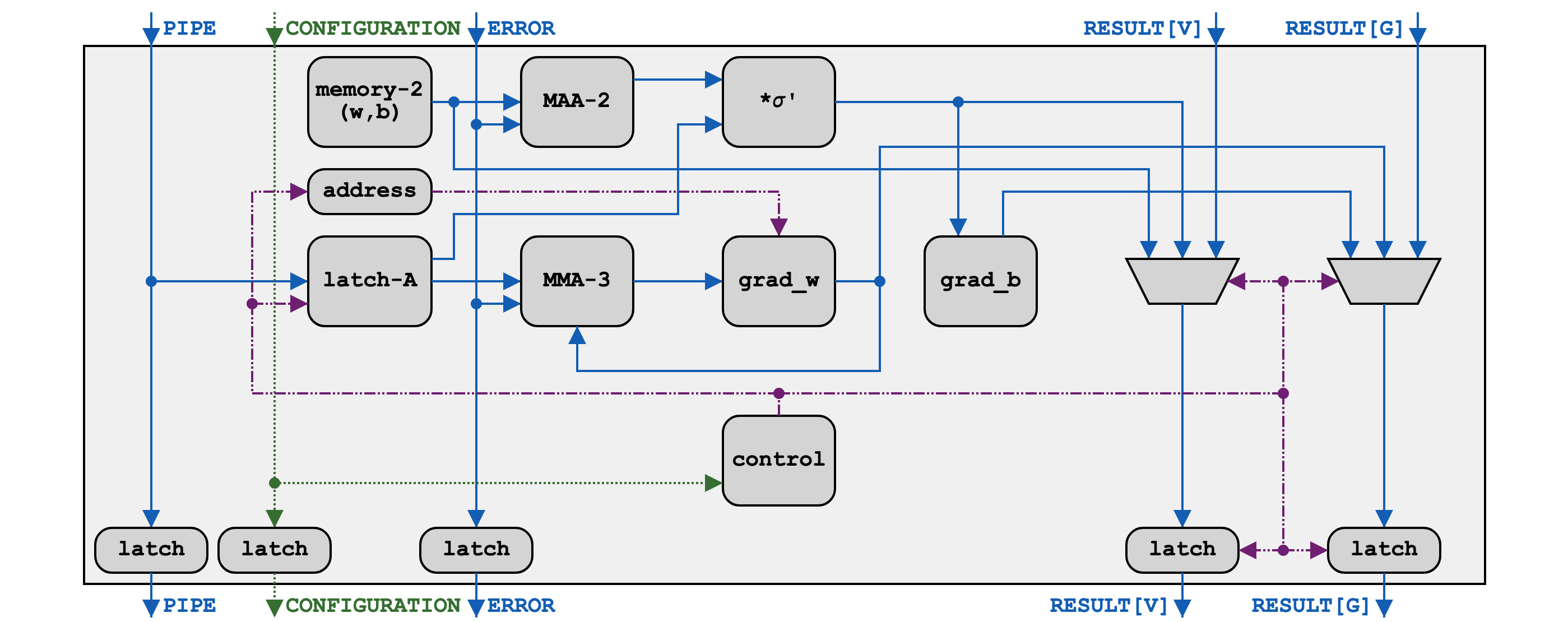} 
    \caption{The internal structure of backward neuron. The top row, placed around \texttt{MAA-2} module, propagates the errors, while middle row, around \texttt{MAA-3}, computes the cost derivatives. The \texttt{RESULT[?]} pipelines serve a dual function: facilitating error propagation during \texttt{learn} state and extracting cost function gradients during \texttt{update} state.}
    \label{fig:bckneuron} 
\end{figure}

The backward neuron is responsible for evaluating Equations~\eqref{eq:bckn_propagation_last}-\eqref{eq:bckn_gradw}, performing both error propagation and gradient computation. Its structure, depicted in Figure~\ref{fig:bckneuron}, is therefore more intricate.

The \texttt{PIPE} stream plays a crucial role in the system. It acts as a pass-through channel sourced from the forward layer and is appropriately delayed by \texttt{delay-1} (see Figure~\ref{fig:design}). This stream carries both the activations and their corresponding derivatives for the next sample being processed. The neuron's control structure captures these values using \texttt{latch-A}, precisely on the clock cycle when the corresponding neuron data is passing through the backward neuron. The \texttt{delay-1} structure ensures that this capture aligns with the two clock cycles allocated for the automatic reset of the remaining parts of the backward neuron. As a result, both the activation and its derivative are reliably latched and ready for use in processing of the next sample.

The same \texttt{CONFIGURATION} that initializes the forward neurons also populates the local memory of the backward neuron, \texttt{memory-2}, with appropriately permuted values\footnote{Permutation is a consequence of the index swap for weights $w$ and can be observed by comparing equations~\eqref{eq:fwdn_stimulus} and~\eqref{eq:bckn_propagation_others}. This permutation can also be observed by comparing indexes in forward and backward neurons on Figure~\ref{fig:math}.}. Once sample errors are received via the \texttt{ERROR} stream, the \texttt{MAA-2} unit evaluates product and summation, the result of which is then multiplied by the precomputed derivative at \texttt{$*\sigma'$}. These two units combined introduce latency $D_{\mathrm{bck}}$, after which their evaluation of Equation~\eqref{eq:bckn_propagation_others} is latched onto the \texttt{RESULT[V]} stream. The system waits until all neurons within the layer complete this process before the computed errors are pushed to the next layer. This mechanism ensures that, analogous to the forward layer, errors are computed and transferred layer to layer.

Simultaneously, gradient computations are performed. Since the $\partial\Gamma/\partial b$ is equivalent to the newly computed error (see Equation~\eqref{eq:bckn_gradb}), it is directly accumulated—once per sample—by the \texttt{grad\_b} component. In contrast, the \texttt{MAA-3} unit evaluates Equation~\eqref{eq:bckn_gradw} for each incoming error estimate. It utilizes the prepared activation and accumulates the computed $\partial\Gamma/\partial w$ in the \texttt{grad\_w} memory.

It is important to note that a neuron in backward layer $l$ computes $\partial\Gamma/\partial b$ for the parameter $b$ in the corresponding forward layer $l$, while $\partial\Gamma/\partial w$ is computed for the parameter $w$ in forward layer $l+1$. See the indices in Figure~\ref{fig:math} for a visual reference. This arrangement is motivated by the fact that the calculation of $\partial\Gamma/\partial w$ depends on error values (i.e., $\partial\Gamma/\partial b$), making it computationally optimal to perform these calculations in the next backward layer. This subsequent layer naturally receives the necessary error information as input due to the structure of error propagation. Shifting these calculations to the following backward layer has two key implications that can be leveraged to optimize the implementation:
\begin{itemize}
  \item The first backward layer computes only $\partial\Gamma/\partial b$ (for the final forward layer).
  \item An additional backward layer is required, which computes only $\partial\Gamma/\partial w$ (for the first forward layer).
\end{itemize}

The design described thus far processes samples, propagates errors, and prepares gradient updates. The remaining components are activated by the control structure upon receiving a network update request. This request is received via the \texttt{CONFIGURATION} stream, targeting one network parameter at a time. Upon receiving the request, the current value of the relevant parameter is latched onto the \texttt{RESULT[V]} pipe, while the corresponding cost gradient is latched onto the \texttt{RESULT[G]} pipe. This is done with $D_{\mathrm{update}}$ latency. Once all neurons in the layer are ready, both pipes are pushed simultaneously. The overal design control repeats this procedure for all network parameters within the layer.

\subsubsection{Update module}

\begin{figure}[htbp] 
    \centering 
    \includegraphics[scale=\imgScale]{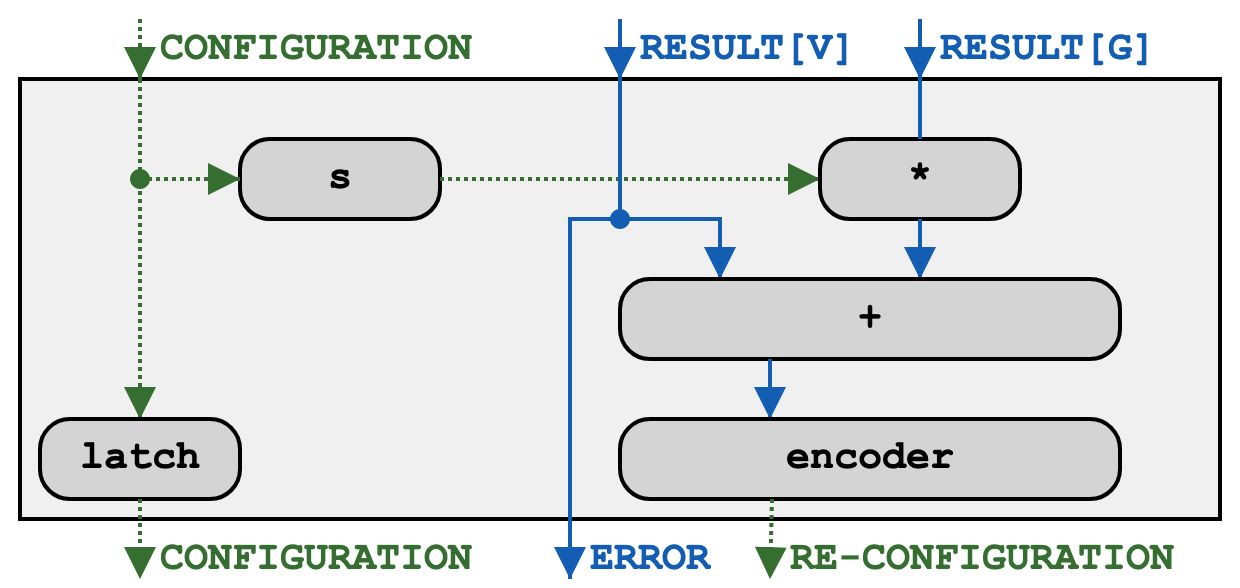} 
    \caption{The internal structure of the update module. During the \texttt{learn} state, the error carried by the \texttt{RESULT} stream is simply passed to the next layer. In the \texttt{update} state, the value and gradient from the \texttt{RESULT} stream are processed to generate the updated parameter value, which is served in form of configuration commands on the \texttt{RE-CONFIGURATION} output.} 
    \label{fig:bckconfig} 
\end{figure}

The streams \texttt{RESULT[V]} and \texttt{RESULT[G]}, generated by the backward neurons, are received by the update module. During the \texttt{learn} state, only \texttt{RESULT[V]} is utilized, as it is simply passed through to the \texttt{ERROR} output stream with layency $D_{\Sigma,\mathrm{error}}$. A functional schematic of this process is shown in Figure~\ref{fig:bckconfig}. In the \texttt{update} state, both streams are used to evaluate equations~\eqref{eq:update_stepb} and~\eqref{eq:update_stepw}. The required scaling parameter, $s$, is set through configuration. The newly computed parameter value is processed by the \texttt{encoder}, which encapsulates it into a properly formatted command and serves it on the \texttt{RE-CONFIGURATION} stream. This stream is then routed back to the design's configuration module, thereby updating the entire design with the new parameter value, thus preparing the design for the next batch.

\subsubsection{Backward layer}

\begin{figure}[htbp] 
    \centering 
    \includegraphics[scale=\imgScale]{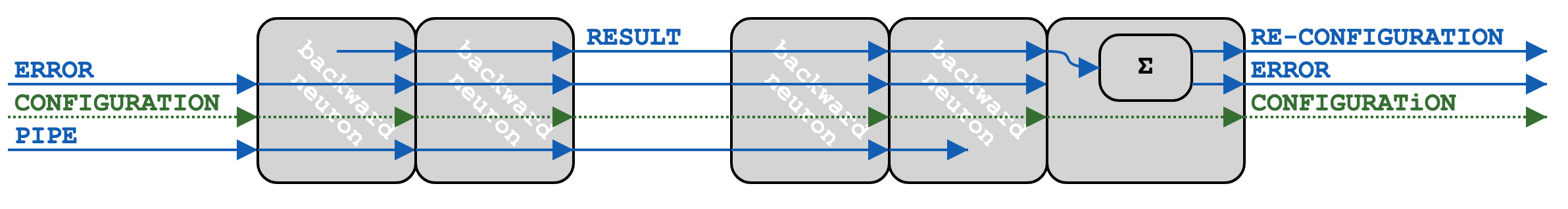} 
    \caption{The internal structure of the backward layer. The same pipeline structure is used as in the forward layer, providing similar benefits. This layer either feeds the next backward layer through the \texttt{ERROR} output or issues system reconfiguration commands, through the \texttt{RE-CONFIGURATION} output, during the update phase.} 
    \label{fig:bcknet} 
\end{figure}

The backward layer is designed in a pipelined manner, leveraging the same principles as the forward layer. Backward neurons are arranged sequentially, with the update structure appended at the end, as illustrated in Figure~\ref{fig:bcknet}.

Since the design mirrors that of the forward layer, it offers the same advantages in terms of parallelism. As in the forward case, it imposes the same restriction on the minimum time interval between sequential samples, thus retaining validity of $p^{\mathrm{tot}}_{\mathrm{sample}}$ of Equation~\eqref{eq:pause}. However, some key differences exist compared to the forward case:

\textbf{Implication 3:} The propagation latency is similar; however, the backward neuron and the update structure exhibit different propagation latency values. Same reasoning as in the forward case can be made, with exception of number of inputs to the backward layer $l$, which equals the number of neurons in the forward layer $l+1$. Thus the total latency induced by a backward layer is given by:
\begin{equation}
    l_{\mathrm{bck}}^l = N_{\mathrm{neurons}} + n^{l+1} + D_{\mathrm{bck}} + D_{\Sigma,\mathrm{error}}
\label{eq:latency_bck}
\end{equation}

\textbf{Implication 4:} Similar timing constraints also apply to period between parameter updates $p_{\mathrm{update}}^l$. Once triggered, a neuron prepares its results on the \texttt{RESULT} stream at time $D_{\mathrm{update}}$. The next update trigger should not occur before the current trigger has produced results for the last neuron, at which point the \texttt{RESULT} pipeline can start shifting. This condition leads to the constraint:
\begin{equation}
    p_{\mathrm{update}}^l \geq n^l.
\label{eq:pause_update}
\end{equation}

\textbf{Implication 5:} Similarly, the latency between issuing an update trigger and the parameters being prepared is $N_{\mathrm{neurons}}+U_{\mathrm{update}}+ D_{\Sigma,\mathrm{update}}$, where the last term represents the delay of the update module calculating new parameter value. Additional time, $D_{\mathrm{config}}$, is required for the parameters to travel from the backward module to the forward module, plus an additional $N_{\mathrm{neurons}}$ for the parameters to reach the last neuron in the forward layer. This results in the total latency from trigger to parameter setting:
\begin{equation}
    L_{\mathrm{update}}^l = 2N_{\mathrm{neurons}}+D_{\mathrm{update}}+ D_{\Sigma,\mathrm{update}} + D_{\mathrm{config}}.
\label{eq:latency_update}
\end{equation}

\subsection{Configurator}

The final module requiring detailed discussion is the \texttt{configurator} block. In \texttt{config} mode, it decodes user commands in the format: layer, neuron, index, parameter, and value. Using this method, all weights and biases can be set accordingly.

A second type of configuration is used to define network-wide parameters. These parameters are implemented as a mapped set of registers, including the number of data inputs, the number of truth inputs, the number of active layers, the number of active neurons within each layer, the delays of delay pipes, the batch size, and the step size. These values are set in the same manner as weights but are targeted by specifying a non-existent layer, with the index field indicating the corresponding register.

An important aspect of this design is that the readout of learned parameters follows the same format and logic. Once triggered, biases and weights are streamed out of the \texttt{LEARNED} port (see Figure~\ref{fig:design}), with one value transmitted per command. This mechanism enables \emph{chaining multiple instances of the entire design}, where the first instance configures the second using the learned parameters. The second design in the chain can then be utilized either for inference or for further training using an additional data stream.

\section{Data Absorption Factor}
\label{sec:absorbtion}

\begin{figure}[htbp] 
    \centering 
    \includegraphics[scale=\imgScale]{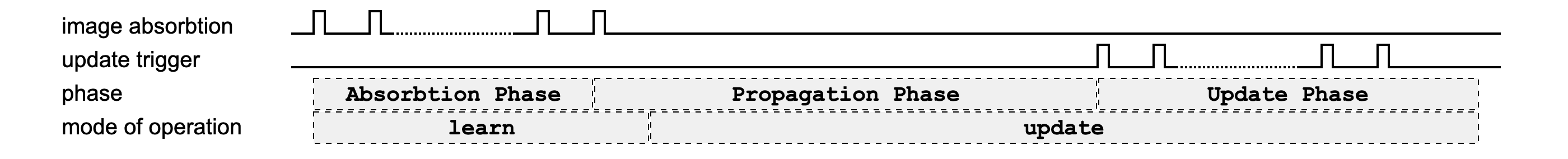} 
    \caption{The timing layout of the learn-update cycle. The \texttt{Absorption Phase} constitutes of the majority of the time the system spends in \texttt{learn} state. During this period, all samples except the last are absorbed by the system. The period required for the system to absorb and fully process the final sample is the \texttt{Propagation Phase}. During this time, the system transitions into \texttt{update} state. Finally the \texttt{Update Phase} covers parameter updates to prepare the system for the next batch.} 
    \label{fig:cycle} 
\end{figure}

The previous section outlined the internal components and their respective functionalities. The key takeaways were five core implications, which now serve as the foundation for characterizing the performance of the design. As stated in the introduction, the primary objective is to maximize data assimilation, enabling the system to utilize the full data set provided by the source. To quantify this capability the concept of data absorption factor is introduced and defined as:
\begin{equation}
    \Upsilon = N_{inputs}/N_{clocks},
\label{eq:factor}
\end{equation}
where $N_{inputs}$ represents the number of data inputs the system can assimilate within $N_{clocks}$ time periods. This value is meaningfully defined when averaged over the entire training. However, since the learn-update cycles are uniform and repeatable, we can evaluate this quantity by examining a single cycle as is depicted in Figure~\ref{fig:cycle}.

Three distinct phases can be identified within each learn-update cycle, each requiring separate analysis:
\begin{itemize} 
    \item \texttt{Absorption Phase}: This phase spans over the majority of the \textit{learn} state. It covers assimilation for all samples, but the last one of the batch. The time spacing between samples is constrained by Equation~\eqref{eq:pause}. Reusing assumption that the system fully utilizes at least one layer, duration of this phase can be expressed:
\begin{equation}
    t_{\mathrm{absorption}} = \max(n_{\mathrm{inputs}}+2, N_{\mathrm{neurons}}+2) \cdot (n_{\mathrm{batch}}-1).
\label{eq:t1}
\end{equation}
    \item \texttt{Propagation Phase}: This phase encompasses the assimilation of the final sample in the batch (while the system is still in the \textit{learn} state) and its propagation through the system (after the system has transitioned into the \textit{update} state). It represents the time required for the final sample to traverse the entire forward network and be fully processed by backward layer $l$, thereby contributing the last necessary information to the gradients of this layer:
\begin{equation}
    t_{\mathrm{propagation}}^l = \sum_{i=1}^{N_{\mathrm{layers}}} l_{\mathrm{fwd}}^i + D_{\mathrm{cost}} + \sum_{i=N_{\mathrm{layers}}}^{l-1} l_{\mathrm{bck}}^i.
\label{eq:t2}   
\end{equation} 
    \item \texttt{Update Phase}: This phase spans the remaining portion of the \textit{update} state, during which update triggers initiate weights reconfiguration, followed by the reconfiguration of biases. In backward layer $l$, each neuron computes $n^{l+1}$ different $\partial\Gamma / \partial w$ terms. Triggers for these must be spaced at least $p_{\mathrm{update}}^l = n^l$ clock cycles apart, thus requiring minimum of $n^{l+1} n^l$ clock periods to trigger all of them. Once this process is completed, an additional $L_{\mathrm{update}}^l$ clock cycles are needed for the reconfiguration of biases. Consequently, the duration of this phase is given by:
\begin{equation}
    t_{\mathrm{update}}^l = n^{l+1} n^l + L_{\mathrm{update}}^l.
\label{eq:t3}
\end{equation}
\end{itemize}
By combining these expressions, the final equation for the data absorption factor is:
\begin{equation}
    \Upsilon = \frac{n_{\mathrm{batch}} \cdot n_{\mathrm{inputs}}}{t_{\mathrm{absorption}} + \max_l (t_{\mathrm{propagation}}^l + t_{\mathrm{update}}^l)}.
\label{eq:factor_final}
\end{equation}
The $\max_l$ function accounts for the fact that the update phase concludes when the last parameter is set in all layers. If the number of neurons, and consequently the number of coefficients, is highly uneven across layers, the final parameter update may not necessarily occur in the last layer but in an earlier layer containing significantly more parameters. This makes the evaluation of this performance metric dependent on the neuron distribution across layers, preventing any general conclusion beyond Equation~\eqref{eq:factor_final}.

\subsection{Evaluation of Applicability}

Although the described system has many degrees of freedom, we can still explore reasonable estimates of the expected $\Upsilon$ values. Equation~\eqref{eq:factor_final} contains several implementation-dependent parameters. While these may vary across different implementations, we can provide a conservative estimate. It will be demonstrated in the following section that the selected set of parameters, listed in Table~\ref{tab:params}, yields a fully functional design. In many implementations, these parameters could even be further optimized.

\begin{table}[h]
    \centering
    \begin{tabular}{|c|c|c|c|}  
        \hline
        $D_{\mathrm{cost}}=3$ & $D_{\mathrm{update}}=2$ & $D_{\mathrm{config}}=2$ & $D_{\Sigma,\mathrm{update}}=3$ \\
        \hline
        $D_{\mathrm{fwd}}=3$  & $D_{\sigma}=3$ & $D_{\mathrm{bck}}=3$  & $D_{\Sigma,\mathrm{error}}=1$ \\
        \hline
    \end{tabular}
    \caption{Conservatively choosen implementation-specific delay parameters.}
    \label{tab:params}
\end{table}

Conservatively assuming that all neurons are utilized, i.e. $n^l = N_{\mathrm{neurons}}$, the system is reduced to one controlled by four parameters: $n_{\mathrm{inputs}}$, $N_{\mathrm{layers}}$, $N_{\mathrm{neurons}}$, and $n_{\mathrm{batch}}$. To further simplify interpretation the network's form factor is defined as:
\begin{equation}
    f = \frac{N_{\mathrm{neurons}}}{n_{\mathrm{inputs}}}.
\label{eq:form_final}
\end{equation}
Intuitively, this metric correlates with different use cases of neural networks. Networks performing simple data reduction typically have a small form factor ($f < 1$), whereas networks used for pattern recognition, classification, and similar tasks tend to have a high form factor ($f > 1$).

\begin{figure}[h]
    \centering
    \begin{tabular}{c c}  
        \includegraphics[width=0.45\textwidth]{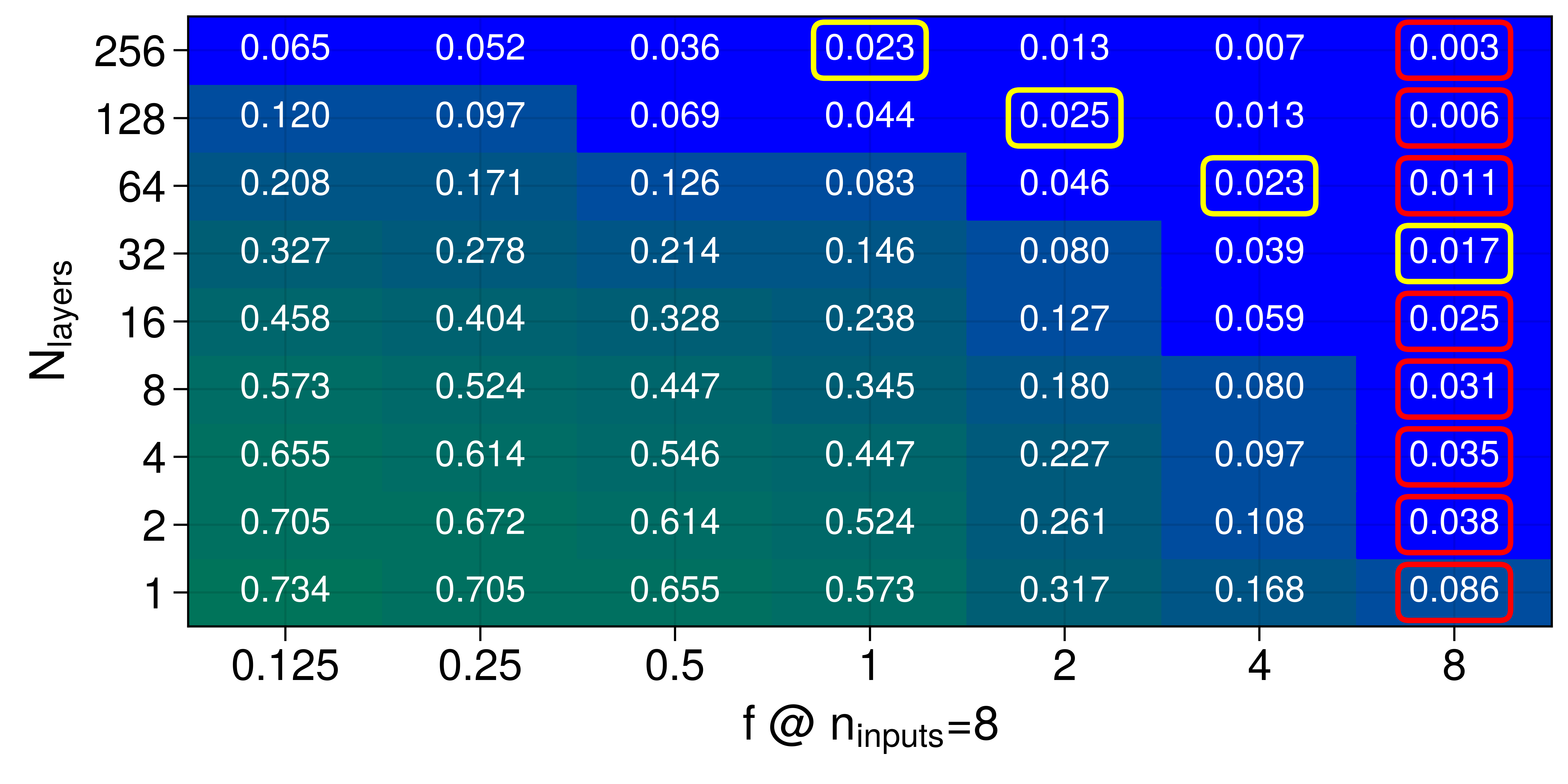} &
        \includegraphics[width=0.45\textwidth]{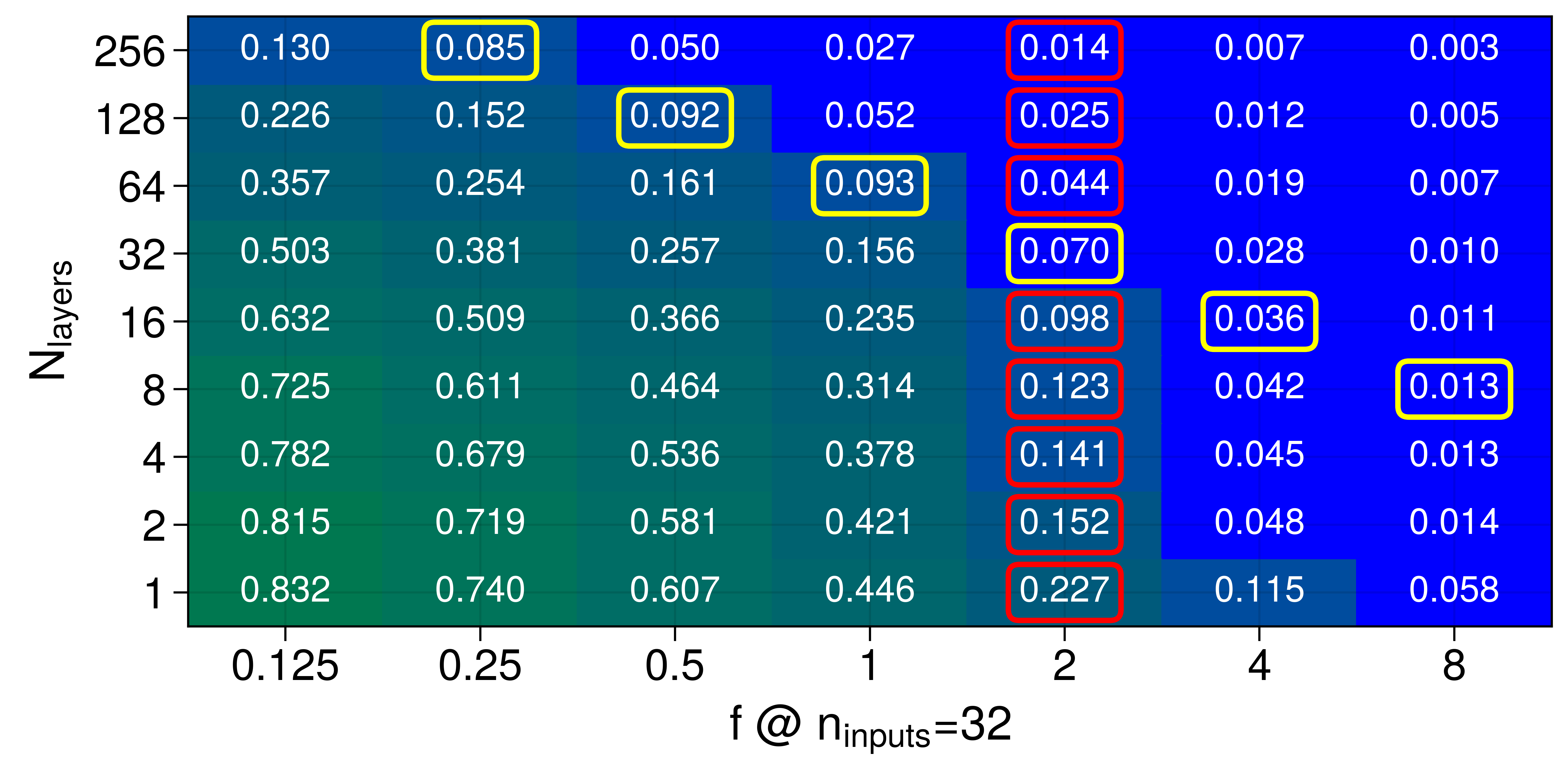} \\
        \includegraphics[width=0.45\textwidth]{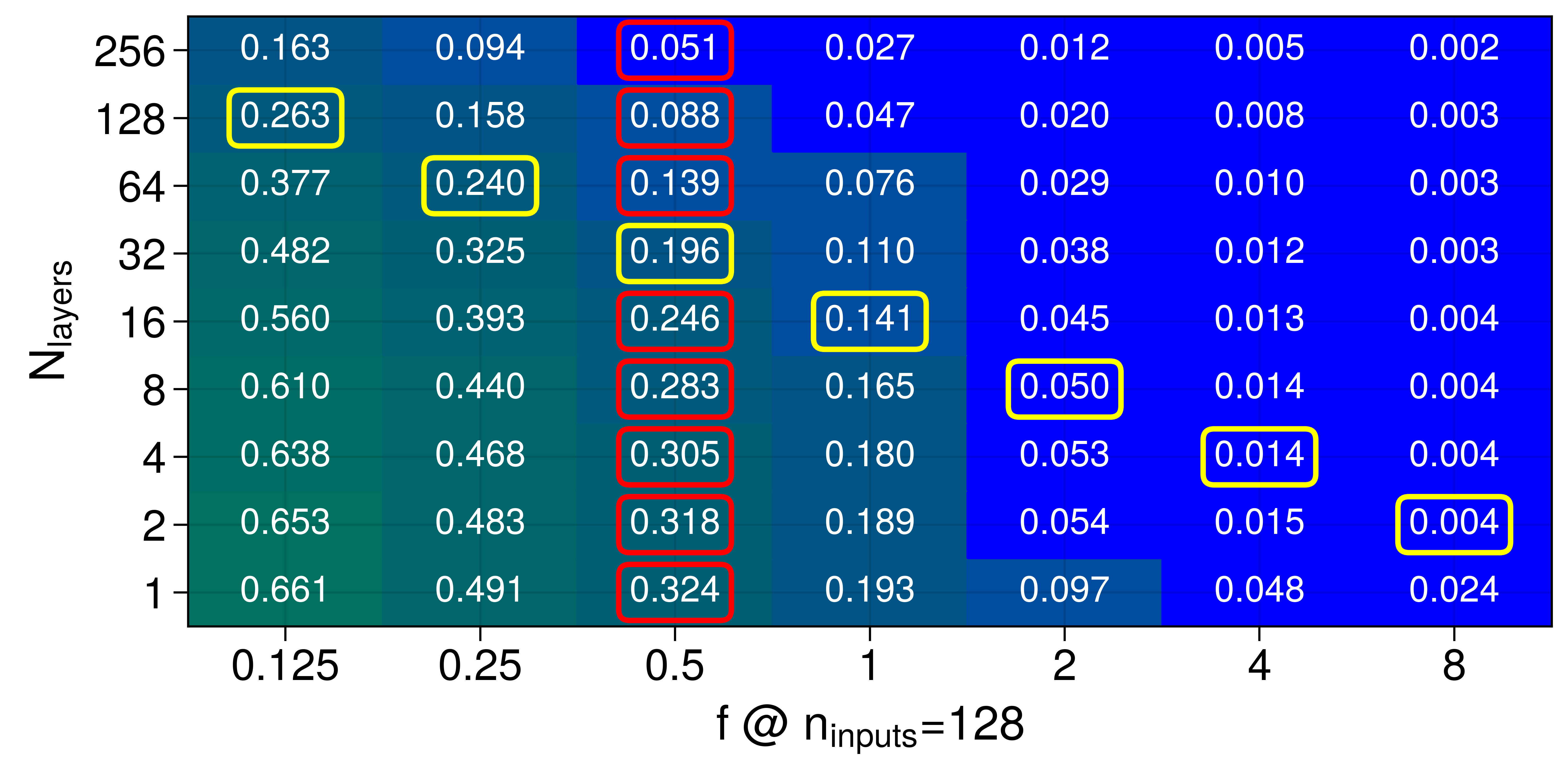} &
        \includegraphics[width=0.45\textwidth]{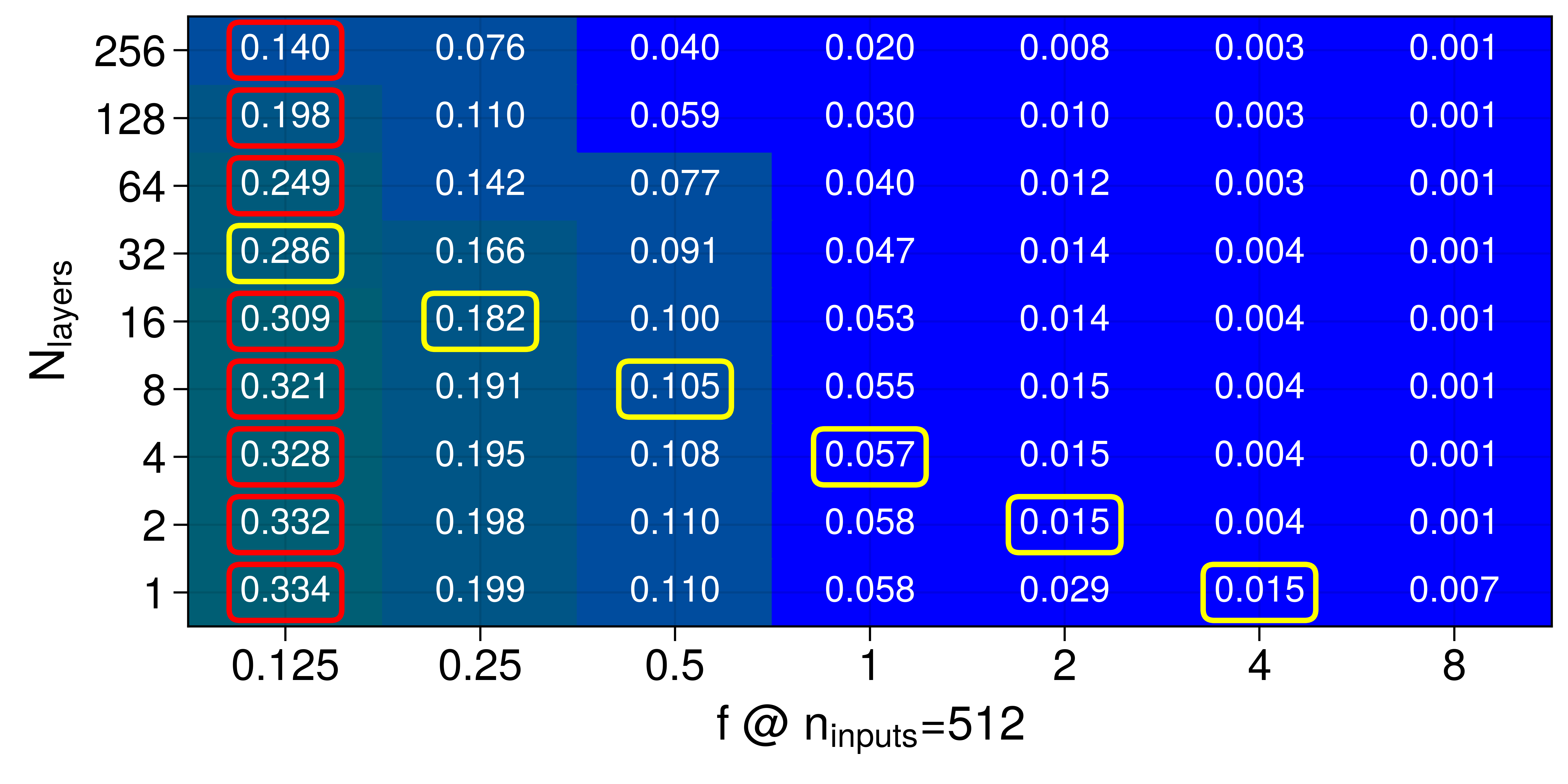} \\
    \end{tabular}
    \caption{Data absorption factor as a function of the number of neuron layers and the network's form factor. The four images represent different input sample sizes used by the network. The colors correspond to the displayed numerical values and help visualize the trend. Yellow boxes mark all designs that have total of $2048$ neurons, while red boxes mark all designs with $64$ neurons per layer.}
    \label{fig:layer_vs_neurons}
\end{figure}

First, $n_{\mathrm{batch}} = 32$ is fixed to explore the absorbtion factor dependence on the three remaining parameters. This is illustrated in Figure~\ref{fig:layer_vs_neurons}. The four plots show the absorption factor as a function of $N_{\mathrm{layers}}$ and the form factor $f$, for different input sample sizes. Note the exponential steps on both axes. Key observations are:
\begin{itemize} 
    \item When  not limited by the number of implementable neurons, the impact of $N_{\mathrm{layers}}$ on performance is relatively minor—especially for larger samples.
     \item When not limited by the number of implementable neurons, the most sensitive parameter is $N_{\mathrm{neurons}}$, which is embedded in the form factor. This sensitivity arises due to the quadratic term in Equation~\eqref{eq:t3}.
    \item When limited by $N_{\mathrm{neurons}}$, the system absorbs data more effectively for larger input samples (equivalent to a smaller network form factor). This trend is visible by following the red boxes in Figure~\ref{fig:layer_vs_neurons}, which correspond to designs with $64$ neurons per layer.
    \item When limited by the total number of implementable neurons, data absorption tends to improve with increasing sample size but remains highly dependent on the network's form factor. This can be observed by following the yellow boxes in Figure~\ref{fig:layer_vs_neurons}, which correspond to designs with a total of $2048$ neurons.
\end{itemize}

\begin{figure}[htbp] 
    \centering 
    \includegraphics[width=0.45\textwidth]{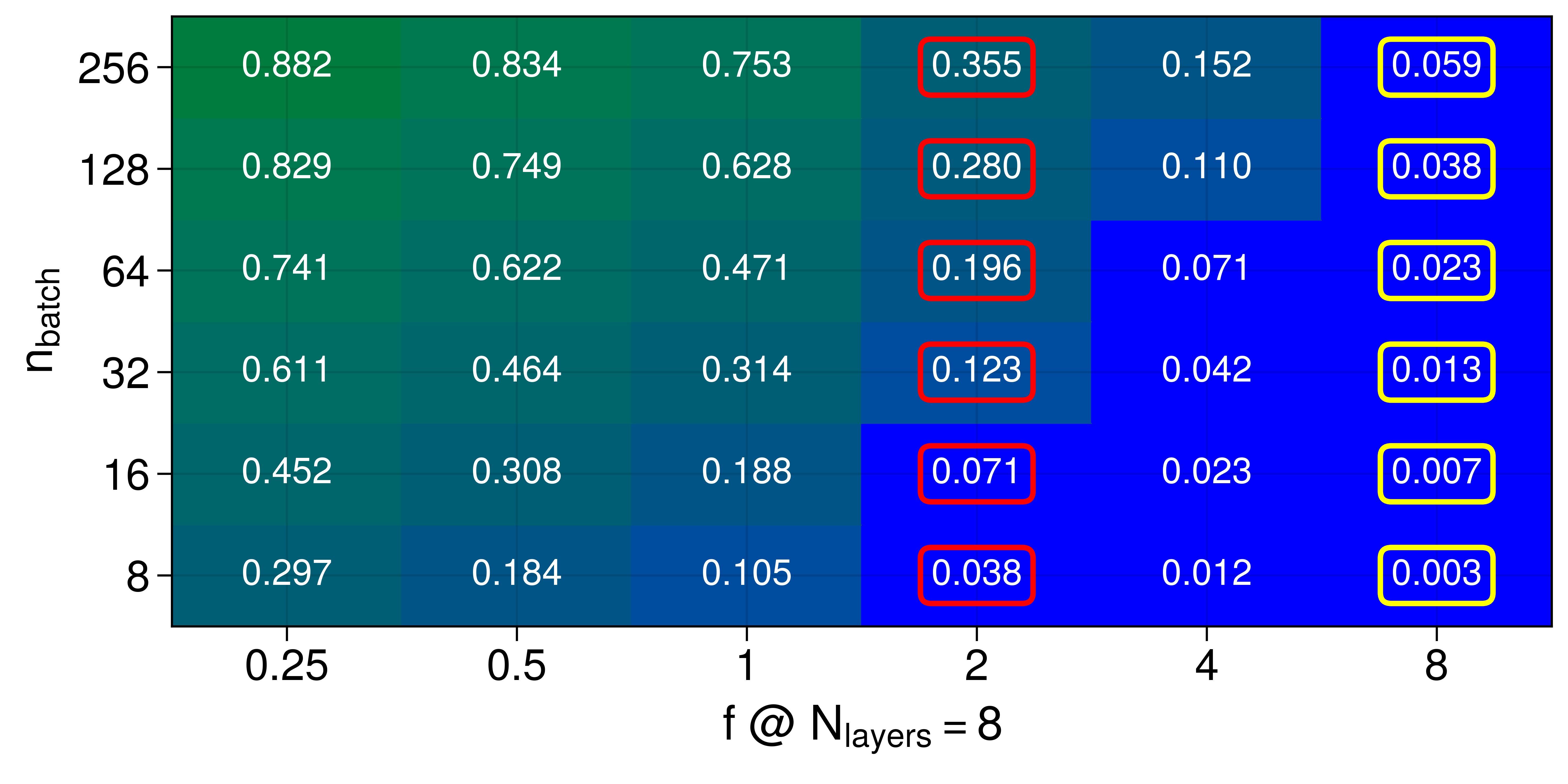} 
    \caption{Data absorption factor as a function of batch size and the network's form factor. The colors correspond to the displayed numerical values and help visualize the significant increase in absorption capability with larger batch sizes. Yellow boxes mark all designs that have total of $2048$ neurons, while red boxes mark all designs with $64$ neurons per layer.} 
    \label{fig:factor_batch} 
\end{figure}

Since the number of layers is the least impactful parameter, let us set $N_{\mathrm{layers}} = 8$ for the rest of this analysis. This allows examination of dependence on $n_{\mathrm{batch}}$, which was previously fixed—see Figure~\ref{fig:factor_batch}. The results demonstrate significant improvement in absorption factor with larger batch sizes.

\begin{figure}[htbp] 
    \centering 
    \includegraphics[width=0.95\textwidth]{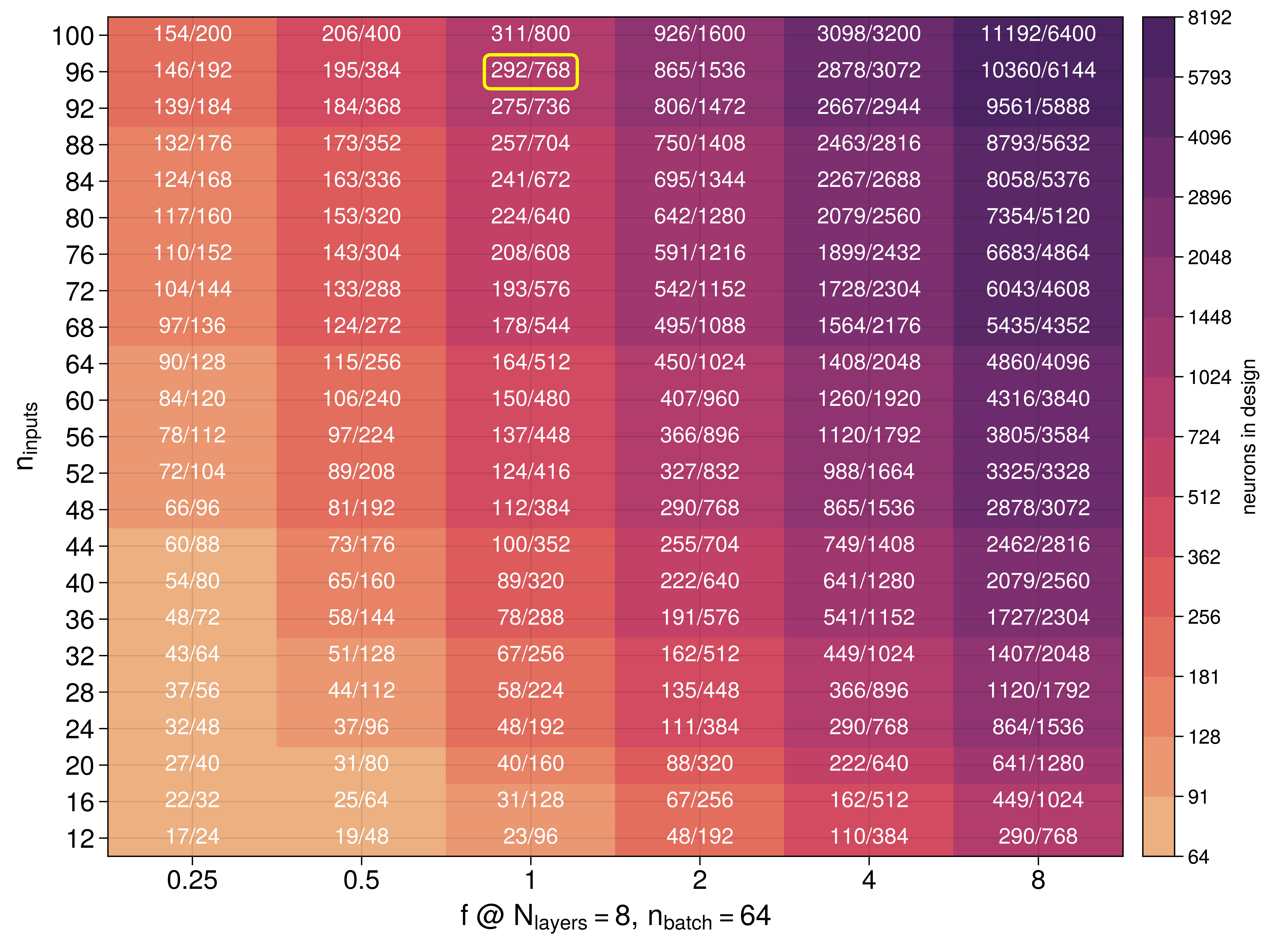} 
    \caption{Applicability chart. This chart shows how the input sample size and the network’s form factor relate to both the number of neurons required for a given design and the period at which samples can be processed. Each field contains two numbers separated by a slash: the number before the slash represents the sample feeding period, while the number after the slash indicates the required number of neurons in the design. The latter also corresponds to the color coding of each field. The chart assumes 8-layer networks with 64 samples per batch. Example (highlighted with a yellow box): If a data source provides samples containing 96 inputs every 300 clock cycles, the design must include at least 768 effective neurons and can achieve a network form factor of up to 1.} 
    \label{fig:applicability} 
\end{figure}

Finally a more comprehensive evaluation of applicability is presented by the applicability chart on Figure~\ref{fig:applicability}. It displays two key quantities: the minimum period between samples $p^{\mathrm{tot}}_{\mathrm{sample}}$ as the first value in the field and the required number of neurons $N_{\mathrm{neurons}}\times N_{\mathrm{layers}}$ as the second value in the field, also corresponding to its color coding. These values are shown as functions of sample input size $n_{\mathrm{inputs}}$ and network form factor $f$. By using the applicability chart, the reader can estimate system performance based on the learning challenge and implementation constraints. As an example, consider an application requiring a network form factor of $4$, constrained by an implementation limit of $2048$ neurons. From the applicability plot, we can foresee such a design being capable of assimilating samples with up to 64 inputs, sustaining this process every $1408$ clock cycles.

The core contribution of this section is the \emph{methodology presented for evaluating system performance}. It is evident that the parameter space is vast, and a detailed study should be replicated for each specific application, taking into account the most significant implementation constraints. However, the method itself demonstrates how to optimize other parameters within the bounds of these limitations. Note also that the given examples were produced for worst-case scenarios and could be improved when more specific application constraints are taken into account.

\section{Design verification}
\label{sec:verification}

To prove the functionality of the presented design, two independent simulations were performed to evaluate its performance on training data. Both simulations implemented fully connected neural networks with six inputs per sample and four layers populated with 64, 64, 16, and 7 neurons, respectively. All layers utilized the PaReLU activation function. Both instances implemented the sum-of-squares cost function described in Section~\ref{sec:cost} and employed floating-point arithmetic using the binary32 format of the IEEE 754 standard~\cite{IEEE754_2019} as the underlying data type.

The first simulation was conducted using a software implementation in \texttt{C}, where a fully connected neural network was implemented. The code did not reflect the structural design described in this document but instead relied on a simple matrix multiplication approach.

The second simulation was implemented in \texttt{VHDL} and performed using Xilinx Vivado tools~\cite{vivado}. It targeted the XCVM1802 System-On-Chip (SoC), a member of Xilinx Versal architecture family~\cite{xilinx_versal_2018}. This architecture was chosen due to its inclusion of \texttt{DSP58} modules, which are DSP engines capable of performing MAA operations on binary32 values. In this implementation the construction of the modules strictly adhered to proposed design. All \texttt{MAA-?} components were implemented using \texttt{DSP58} modules, and all storage components utilized \texttt{RAMB18} modules. Omitting further technical details, this implementation exhibits the exact timing characteristics outlined in Table~\ref{tab:params}. It is important to emphasize that no efforts were made to optimize performance or resource utilization. The primary objective was solely to verify the design's functionality.

With both simulations briefly described, we now focus on the dataset used—HAR70+\cite{har70}. This is a classification dataset with six features and seven possible classifications. The first of eighteen subsamples was used and split into a training set containing $100,000$ samples and a test set with $3,860$ samples.

\begin{figure}[htbp] 
    \centering 
    \includegraphics[width=0.9\textwidth]{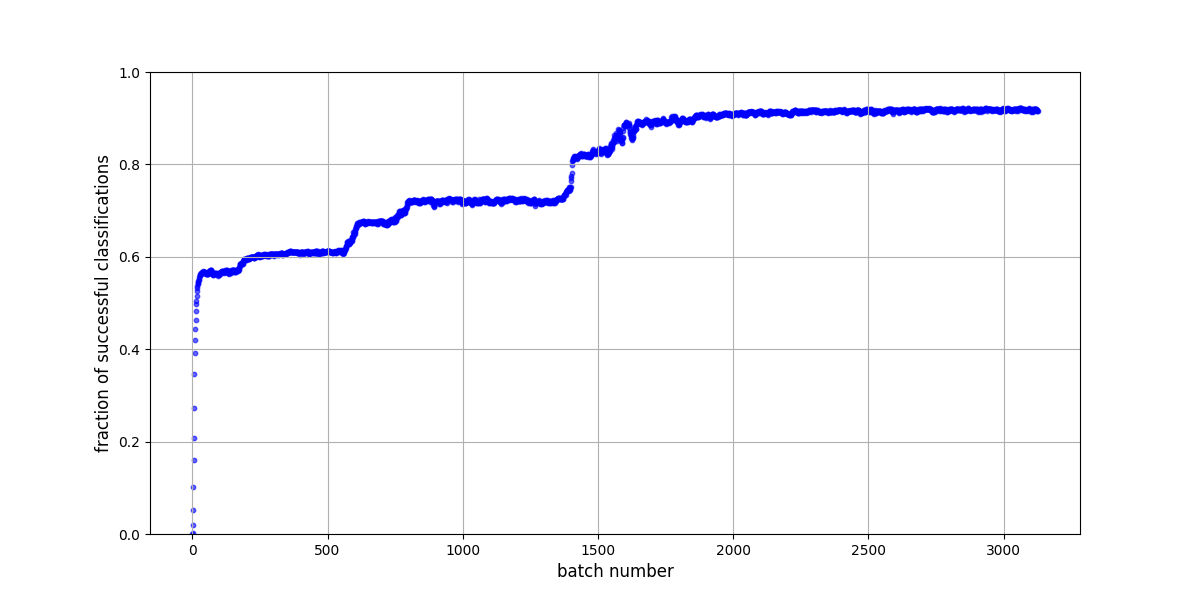} 
    \caption{Classification success during the learning process. The training lasted for two epochs, meaning each training sample was used twice. The design demonstrates the ability to learn the sample features, yielding identical results in both simulations.} 
    \label{fig:learning} 
\end{figure}

The described dataset was then processed using both simulations. Both shared the same initial network parameters, which were randomly initialized using a Gaussian distribution. Training was conducted with $n_{\mathrm{batch}} = 64$ and a step size of $s = 1 \cdot 10^{-5}$ for two epochs. The fraction of correctly classified test samples as a function of batch number is illustrated in Figure~\ref{fig:learning}.

The result demonstrates that the proposed design converges with high classification success. It should be noted that the hyperparameters of the neural network were not optimized, and there is no claim that the presented results are optimal. The key takeaway is that both the simplistic software implementation and the detailed firmware implementation—strictly adhering to the proposed design—produced \emph{identical results}. Final trained parameters were compared between the two simulations and were found \emph{identical in all bits of binary32 format for all network parameters}. This serves as a verification of the design, demonstrating that the \emph{proposed architecture does not compromise data quality or learning capability}.

\subsection{Performance and Resource Estimation}
\label{sec:estimation}

An additional benefit of the \texttt{VHDL} implementation is the ability to perform detailed analysis of timing performance and resource utilization. The implementation was carried through to bitstream generation, with a design clock constraint of $160~\mathrm{MHz}$. The place and route report shows a worst setup slack of $1.46~\mathrm{ns}$ and a worst hold slack of $0.01~\mathrm{ns}$\footnote{Setup slack is defined as the margin by which a signal arrives before the latest allowable time defined by the setup constraints of circuit components. Similarly, hold slack is the margin by which a signal arrives after the earliest allowable time defined by the components hold constraints.}, indicating a comfortable margin in timing performance. This confirms the system can operate reliably at $160~\mathrm{MHz}$.

\begin{table}[h]
  \centering
  \begin{tabular}{|c|c|c|}
    \hline
    Design Structure & \texttt{DSP58} Count & Comment \\
    \hline
    Forward network & 256 & $=64~\mathrm{neurons}\times 4~\mathrm{layers}\times 1~\mathrm{DSP58/neuron}$ \\
    Backward network & 1216 & $=64~\mathrm{neurons}\times 5~\mathrm{layers}\times 3~\mathrm{DSP58/neuron}$ \\
    Cost structure & 1 & simple sum-of-squares cost function \\
    Update module & 5 & $=5~\mathrm{layers}\times 1~\mathrm{DSP58/layer}$ \\
    \hline
    \hline
    Design Structure & \texttt{RAMB18} Count & Comment \\
    \hline
    Forward network & 256 & $=64~\mathrm{neurons}\times 4~\mathrm{layers}\times 1~\mathrm{RAMB18/neuron}$ \\
    Backward network & 640 & $=64~\mathrm{neurons}\times 5~\mathrm{layers}\times 2~\mathrm{RAMB18/neuron}$ \\
    Layer delay pipes & 10 & $=4+2+2+1+1$ for individual delay pipes \\
    Truth delay pipe & 2 &  \\
    \hline
  \end{tabular}
  \caption{Place \& route utilization report of \texttt{DSP58} and \texttt{RAMB18} usage for the fully implemented design.}
  \label{tab:usage_dsp}
\end{table}

The most relevant results from the utilization report are presented in Table~\ref{tab:usage_dsp}. The implementation used $1222$ \texttt{DSP58} blocks and $908$ \texttt{RAMB18} blocks, corresponding to $62\%$ and $47\%$ of the available resources on the XCVM1802 SoC, respectively. The ratio of available \texttt{DSP58} to \texttt{RAMB18} blocks is typically consistent across the same device family, so \texttt{DSP58} blocks are expected to be the primary limiting factor for design size. On average, the implementation required $4.8$ \texttt{DSP58} blocks per effective neuron\footnote{An effective neuron is defined as the ensemble of all components required to perform both forward and backward propagation for a single neuron.}, a metric which estimates the feasible network scale within a single chip.

As of writing, the Xilinx Versal architecture offers up to approximately $14k$ \texttt{DSP58} blocks per chip, suggesting that around $3k$ effective neurons could be implemented on a single device. According to Figure~\ref{fig:applicability}, this enables networks with up to $8$ layers, each with $368$ neurons, capable of processing samples with $92$ floating-point values every $2,667$ clock cycles.

\section{Discussion}
\label{sec:dicsussion}

This paper presented a novel approach for training fully connected neural networks directly in digital hardware, with a specific emphasis on maximizing data absorption throughput. It was demonstrated that in applications where data assimilation presents a significant bottleneck, one does not necessarily need to sacrifice model accuracy: \emph{the performance of the proposed design equals that of conventional software-based methods}. Moreover, the evaluation methodology introduced here provides a predictive framework for assessing the system’s applicability across a range of scenarios. All estimations were made conservatively, suggesting that with detailed optimization tailored to a specific use case, further performance improvements are feasible.

The main strengths of the proposed design can be summarized as follows:
\begin{itemize}
  \item \textbf{Full precision is preserved}, in contrast to many accelerator designs that rely on aggressive compression, quantization or pruning techniques.
  \item \textbf{High data absorption speed}, achieved through architectural optimization focused specifically on real-time input processing.
  \item \textbf{Self-contained deployment}, requiring only a digital clock and an input data stream—ideal for embedding directly into detector front-end systems.
\end{itemize}

Given these features, the system can function as a self-contained processing block within detector electronics, enabling feature extraction and learning before any external filtering is applied—thus fulfilling the central goal of this research. However, this is only possible in cases where the ground truth is available prior to filtering. It should be emphasized that this is not always the case. In context of HEP the system will need to be applied in a specialized manner as the primary objective is unsupervised learning.

The simplest applications in such environments would employ autoencoders, where the data itself serves as the ground truth and the network architecture can be appropriately constrained. Another possibility is the use of generative adversarial networks, which introduces the additional challenge of managing the cost structures for both networks, as they must exchange information. Alternatively, the design could be implemented in an “onion-like” fashion: stacking forward networks before the overall cost structure, followed by a mirrored stack of backward networks. This topology approaches the structure of normalizing flows. All of these approaches introduce extra challenges, such as increased structural overhead and the need for local random number generators which replace the ground truth in unsupervised learning. However, a detailed exploration of these advanced architectures lies beyond the scope of this research, which—as stated in the introduction—focuses on understanding the complexity, trade-offs, and performance of the main structural component.

To contextualize the performance results, we compare the system’s processing capacity with conventional metrics such as Tera-Operations-Per-Second (TOPS). Based on the estimate from the previous section, a configuration with $3k$ effective neurons, each performing three MAA operations in parallel, yields $9k$ MAA operations per clock cycle. At a esimated clock speed of $160~\mathrm{MHz}$, this corresponds to approximately $1.4~\mathrm{TOPS}$. For comparison, the NVIDIA Jetson Nano—a widely used edge device—achieves about $0.125~\mathrm{TOPS}$ at floating-point precision and has been studied extensively for low-power inference tasks in~\cite{SWAMINATHAN2025906, technologies12060081}.

However, raw computational power is not the sole figure of merit. Overall system performance—including latency and data throughput—is often more relevant. A useful comparative metric is the total number of clock cycles required to process a single sample, $p_{\mathrm{sample}}^{\mathrm{tot}}$. An informative review of ML applications in high-energy physics can be found in~\cite{10.3389/fdata.2020.598927}, which reports typical inference periods in the range of $22-55$ clock cycles per sample. According to Figure~\ref{fig:applicability}, the presented system can achieve comparable periods for training with inputs comprising fewer than 16 float32 values, even under single-chip constraints.

Another reference is provided by the survey in~\cite{9043731}, which catalogs performance gains across diverse algorithm–hardware pairings. While this offers a broad overview, direct comparison is difficult due to the inherent differences between inference and training workloads, as well as the variability in accuracy–latency trade-offs across studies. This further underscores the uniqueness of the current work, which offers precise, application-driven predictions for training performance - an area still largely unexplored.

In HEP, the most relevant performance metrics are network size and sample intake period. These metrics can be used to estimate how many of the produced experimental repetitions can be monitored without data loss. The LHC, taken as a reference, features $3,564$ possible proton collision slots, with collisions occurring at a frequency of $40~\mathrm{MHz}$. According to the results presented in Section~\ref{sec:estimation}, the largest currently feasible single-chip design can absorb $92$ floating-point numbers every $2,667$ clock cycles at an operating frequency of $160~\mathrm{MHz}$. This corresponds to a sample intake period of $p_{\mathrm{sample}}^{\mathrm{tot}} \sim 667$ in units of the proton collision clock. Consequently, such a system could assimilate data from $5$ choosen LHC proton clouds without any dead-time. It is important to note that this estimate is based on the largest possible network and therefore represents a conservative lower bound; significantly larger ensembles could be supported by employing smaller networks.

Several recent ML applications illustrate the relevant network sizes in HEP. High-level jet simulation has been successfully improoved using networks of approximately 512 neurons\footnote{Applying the reasoning from the previous paragraph, while maintaining a form factor requirement of $f=4$, yields an estimated capability of monitoring approximately $90$ LHC proton clouds without any dead-time.} across 3 layers~\cite{Qu_2020}. Likewise, studies focused on low-latency inference have explored architectures with $3-5$ layers and $32-128$ neurons per layer, totaling around $600$ neurons~\cite{Duarte_2018}. These examples affirm that many meaningful HEP tasks are already within the practical deployment range of the proposed architecture.

In summary, this work demonstrates that real-time, in-situ and unbiased training of neural networks is not only feasible but also practically impactful. By targeting data-rich environments and focusing on throughput, this design enables a new class of on-detector intelligence that operates before traditional filtering—a direction previously unexplored.

\section{Outlook}
\label{sec:implementation}

This study has primarily focused on establishing the design principles of a novel digital hardware architecture for neural network training and evaluating its intrinsic limitations. By characterizing key performance bounds, it offers a reliable framework for estimating the design's applicability across a wide range of scenarios. Given its potential in HEP, the next logical step is to apply the architecture to a specific use case and assess its impact.

A crucial first task in this direction is to conduct a detailed physics-driven case study. The performance estimates derived here provide constraints on feasible network sizes, which can guide the formulation of a concrete physics application. In particular, implementing one of the unsupervised learning approaches mentioned in the introduction—built upon the neural network structure explored in this work—would provide a compelling demonstration of the architecture’s relevance.

In parallel, the process of refining application-specific constraints will naturally inform the final implementation strategy, enabling optimization of the hardware design. The architecture is implementable with various technologies, including GPU arrays, FPGAs, and custom ASICs. While a full comparison of these technologies is beyond the scope of this paper, it is worth emphasizing the design’s flexibility by outlining a few key optimization pathways:

\begin{itemize}
  \item \textbf{Scalability}: The pipeline-oriented, minimally interconnected structure of the architecture lends itself naturally to scaling. For instance, a larger network can be distributed across multiple FPGA devices with minimal interdependency.
  
  \item \textbf{Forward/backward pipelining}: The architecture supports partial parallelization of neuron operations. Neurons can be grouped into “super neurons” capable of processing multiple inputs per clock cycle and computing multiple activations in parallel. This strategy effectively halves the absorption time, $\mathrm{t_{absorption}}$, and simultaneously reduces the latency of the forward and backward passes.
  
  \item \textbf{Memory optimization}: Enhancing the \texttt{memory-\textit{?}} units with multi-port read/write capabilities would enable parallel parameter updates. For example, updating two parameters concurrently would alleviate the quadratic bottleneck identified in Equation~\eqref{eq:t3}, significantly benefiting large-scale networks.
  
  \item \textbf{Time-multiplexing}: As shown in Section~\ref{sec:dicsussion}, the system exceeds the required data absorption speed for learning from individual proton clouds. This opens up two promising avenues: (i) learning from aggregated clouds for improved generalization, or (ii) time-multiplexing the architecture to train multiple networks in parallel with a single device. Given the capacity of modern block RAMs, this would allow for simultaneous learning on multiple data streams—ideal for systematic studies.
\end{itemize}

While real-world deployment will require further tailoring to specific experimental constraints, the core architecture provides ample flexibility and optimization potential for a range of platforms.

The most immediate implementation candidate is FPGA-based hardware, where this architecture is projected to support up to $3,000$ neurons per chip. This figure represents a key constraint for practical deployments and serves as a benchmark for application planning. It should be noted, however, that the presented FPGA-based demonstration design relied on floating-point numbers, which can be efficiently processed by the \texttt{DSP58} blocks available in the Xilinx Versal product family. Other Xilinx product families lack this capability, which currently limits the range of compatible devices. Conversely, this raises the prospect of fixed-point implementations and quantization of the presented design, which could be a promising direction for future exploration.

Nevertheless, the architecture is not confined to FPGA technology. ASIC implementations, in particular, offer compelling opportunities for further performance gains, both in clock speed and data throughput. In such a setting, the design could be tightly optimized to meet the most demanding scientific or industrial requirements.

With all major performance metrics modeled, this work lays a solid foundation for near-term implementation efforts. It holds particular promise for low-latency, high-throughput environments such as HEP experiments, where early-stage autonomous learning may become essential to managing future data complexity.

\section*{Abbreviations}

The following abbreviations are used in this manuscript:
\\

\noindent 
\begin{tabular}{@{}ll}
ASIC & Application Specific Integrated Circuit \\
CPU & Central Processing Unit \\
FIFO & First-In-First-Out \\
FPGA & Field Programmable Gate Array \\
GPU & Graphics Processing Unit \\
HEP & High-Energy Physics \\
LHC & Large Hadron Collider \\
MAA & Multiply-And-Accumulate \\
ML & Machine Learning \\
SoC & System-on-Chip \\
TPU & Tensor Processing Unit \\
TOPS & Tera Operations Per Second
\end{tabular}

\section*{Funding}

The author acknowledges the financial support from the Slovenian Research and Innovation Agency (ARIS P1-0135).

\acknowledgments

I would like to thank dr. Andrej Seljak for valuable discussions, insightful opinions, and constructive debates that contributed to the development of this work.


\bibliographystyle{unsrt}  
\bibliography{references}  

\begin{thebibliography}{10}

\bibitem{LE2008}
Lyndon Evans and Philip Bryant.
\newblock Lhc machine.
\newblock {\em Journal of Instrumentation}, 3(08):S08001, aug 2008.

\bibitem{mlrates}
Philip Harris, Erik Katsavounidis, William McCormack, Dylan Rankin, Yunyang
  Feng, Abhijith Gandrakota, Christian Herwig, Burt Holzman, Kevin Pedro, Nga
  Tran, Tingjun Yang, Jennifer Ngadiuba, Michael Coughlin, Scott Hauck,
  Shih-Chieh Hsu, Elham Khoda, Deming Chen, Mark Neubauer, Joel Duarte, and Mia
  Liu.
\newblock Physics community needs, tools, and resources for machine learning,
  03 2022.

\bibitem{AtlasTrigger}
The~ATLAS collaboration.
\newblock Operation of the atlas trigger system in run 2.
\newblock {\em Journal of Instrumentation}, 15(10):P10004, oct 2020.

\bibitem{GUPTA20211}
Neha Gupta.
\newblock {\em Hardware Accelerator Systems for Artificial Intelligence and
  Machine Learning}, volume 122 of {\em Advances in Computers}.
\newblock Elsevier, 2021.

\bibitem{Burhanuddin_2023}
M.A. Burhanuddin.
\newblock Efficient hardware acceleration techniques for deep learning on edge
  devices: A comprehensive performance analysis.
\newblock {\em KHWARIZMIA}, 2023:1--10, 08 2023.

\bibitem{Tao_2020_CVPR_Workshops}
Yudong Tao, Rui Ma, Mei-Ling Shyu, and Shu-Ching Chen.
\newblock Challenges in energy-efficient deep neural network training with
  fpga.
\newblock In {\em Proceedings of the IEEE/CVF Conference on Computer Vision and
  Pattern Recognition (CVPR) Workshops}, June 2020.

\bibitem{MOOLCHANDANI2021101887}
Diksha Moolchandani, Anshul Kumar, and Smruti~R. Sarangi.
\newblock Accelerating cnn inference on asics: A survey.
\newblock {\em Journal of Systems Architecture}, 113:101887, 2021.

\bibitem{MACHUPALLI2022104441}
Raju Machupalli, Masum Hossain, and Mrinal Mandal.
\newblock Review of asic accelerators for deep neural network.
\newblock {\em Microprocessors and Microsystems}, 89:104441, 2022.

\bibitem{CHEN2020264}
Yiran Chen, Yuan Xie, Linghao Song, Fan Chen, and Tianqi Tang.
\newblock A survey of accelerator architectures for deep neural networks.
\newblock {\em Engineering}, 6(3):264--274, 2020.

\bibitem{yan2024surveyfpgabasedacceleratorml}
Feng Yan, Andreas Koch, and Oliver Sinnen.
\newblock A survey on fpga-based accelerator for ml models, 2024.

\bibitem{10940371}
Gaurav Tiwari, Sangeeta Nakhate, Alok Pathak, Abhinandan Jain, and Shardul
  Penurkar.
\newblock Hardware accelerators for deep learning applications, 2025.

\bibitem{fi12070113}
Maurizio Capra, Beatrice Bussolino, Alberto Marchisio, Muhammad Shafique, Guido
  Masera, and Maurizio Martina.
\newblock An updated survey of efficient hardware architectures for
  accelerating deep convolutional neural networks.
\newblock {\em Future Internet}, 12(7), 2020.

\bibitem{10.1145/3174243.3174258}
Duncan~J.M Moss, Srivatsan Krishnan, Eriko Nurvitadhi, Piotr Ratuszniak, Chris
  Johnson, Jaewoong Sim, Asit Mishra, Debbie Marr, Suchit Subhaschandra, and
  Philip~H.W. Leong.
\newblock A customizable matrix multiplication framework for the intel harpv2
  xeon+fpga platform: A deep learning case study.
\newblock In {\em Proceedings of the 2018 ACM/SIGDA International Symposium on
  Field-Programmable Gate Arrays}, FPGA '18, page 107–116, New York, NY, USA,
  2018. Association for Computing Machinery.

\bibitem{10.1145/3307650.3322237}
Mohsen Imani, Saransh Gupta, Yeseong Kim, and Tajana Rosing.
\newblock Floatpim: in-memory acceleration of deep neural network training with
  high precision.
\newblock In {\em Proceedings of the 46th International Symposium on Computer
  Architecture}, ISCA '19, page 802–815, New York, NY, USA, 2019. Association
  for Computing Machinery.

\bibitem{JMLR:v24:22-1208}
Brian~R. Bartoldson, Bhavya Kailkhura, and Davis Blalock.
\newblock Compute-efficient deep learning: Algorithmic trends and
  opportunities.
\newblock {\em Journal of Machine Learning Research}, 24(122):1--77, 2023.

\bibitem{10.3389/fdata.2022.787421}
Allison~McCarn Deiana, Nhan Tran, Joshua Agar, Michaela Blott, Giuseppe
  Di~Guglielmo, Javier Duarte, Philip Harris, Scott Hauck, Mia Liu, Mark~S.
  Neubauer, Jennifer Ngadiuba, Seda Ogrenci-Memik, Maurizio Pierini, Thea
  Aarrestad, Steffen Bähr, Jürgen Becker, Anne-Sophie Berthold, Richard~J.
  Bonventre, Tomás~E. Müller~Bravo, Markus Diefenthaler, Zhen Dong, Nick
  Fritzsche, Amir Gholami, Ekaterina Govorkova, Dongning Guo, Kyle~J.
  Hazelwood, Christian Herwig, Babar Khan, Sehoon Kim, Thomas Klijnsma, Yaling
  Liu, Kin~Ho Lo, Tri Nguyen, Gianantonio Pezzullo, Seyedramin Rasoulinezhad,
  Ryan~A. Rivera, Kate Scholberg, Justin Selig, Sougata Sen, Dmitri Strukov,
  William Tang, Savannah Thais, Kai~Lukas Unger, Ricardo Vilalta, Belina von
  Krosigk, Shen Wang, and Thomas~K. Warburton.
\newblock Applications and techniques for fast machine learning in science.
\newblock {\em Frontiers in Big Data}, Volume 5 - 2022, 2022.

\bibitem{hepMLinfrastructure}
M.~Migliorini, R.~Castellotti, L.~Canali, and M.~Zanetti.
\newblock Machine learning pipelines with modern big data tools for high energy
  physics.
\newblock {\em Computing and Software for Big Science}, 4(1):8, 2020.

\bibitem{QIAN2021165527}
Zhen Qian, Vladislav Belavin, Vasily Bokov, Riccardo Brugnera, Alessandro
  Compagnucci, Arsenii Gavrikov, Alberto Garfagnini, Maxim Gonchar, Leyla
  Khatbullina, Ziyuan Li, Wuming Luo, Yury Malyshkin, Samuele Piccinelli, Ivan
  Provilkov, Fedor Ratnikov, Dmitry Selivanov, Konstantin Treskov, Andrey
  Ustyuzhanin, Francesco Vidaich, Zhengyun You, Yumei Zhang, Jiang Zhu, and
  Francesco Manzali.
\newblock Vertex and energy reconstruction in juno with machine learning
  methods.
\newblock {\em Nuclear Instruments and Methods in Physics Research Section A:
  Accelerators, Spectrometers, Detectors and Associated Equipment},
  1010:165527, 2021.

\bibitem{duarte2022fastmlsciencebenchmarksaccelerating}
Javier Duarte, Nhan Tran, Ben Hawks, Christian Herwig, Jules Muhizi, Shvetank
  Prakash, and Vijay~Janapa Reddi.
\newblock Fastml science benchmarks: Accelerating real-time scientific edge
  machine learning, 2022.

\bibitem{bartoldus2022innovationstriggerdataacquisition}
Rainer Bartoldus, Catrin Bernius, and David~W. Miller.
\newblock Innovations in trigger and data acquisition systems for
  next-generation physics facilities, 2022.

\bibitem{201925}
Real-time data processing in the alice high level trigger at the lhc.
\newblock {\em Computer Physics Communications}, 242:25--48, 2019.

\bibitem{8424231}
N.~M. Truong, M.~Aoki, Y.~Igarashi, M.~Saito, S.~Ito, D.~Nagao, Y.~Nakatsugawa,
  H.~Natori, Y.~Seiya, N.~Teshima, and K.~Yamamoto.
\newblock Real-time lossless compression of waveforms using an fpga.
\newblock {\em IEEE Transactions on Nuclear Science}, 65(9):2650--2656, 2018.

\bibitem{AXIOTIS}
Konstantinos AXIOTIS.
\newblock Studies on fpga-based solutions for the atlas phase ii trigger and
  data acquisition system.
\newblock {\em Doctoral Thesis}, 2025.

\bibitem{doi:10.1142/9789811234033_0012}
Javier Duarte and Jean-Roch Vlimant.
\newblock {\em Graph Neural Networks for Particle Tracking and Reconstruction},
  chapter Chapter 12, pages 387--436.

\bibitem{bileska2025designfpgaimplementationwombat}
Mila Bileska.
\newblock Design and fpga implementation of wombat: A deep neural network
  level-1 trigger system for jet substructure identification and boosted
  $h\rightarrow b\bar{b}$ tagging at the cms experiment, 2025.

\bibitem{inferenceFPGAhep}
Javier Duarte, Philip Harris, Scott Hauck, Burt Holzman, Shih-Chieh Hsu, Sergo
  Jindariani, Suffian Khan, Benjamin Kreis, Brian Lee, Mia Liu, Vladimir Lon{\v
  c}ar, Jennifer Ngadiuba, Kevin Pedro, Brandon Perez, Maurizio Pierini, Dylan
  Rankin, Nhan Tran, Matthew Trahms, Aristeidis Tsaris, Colin Versteeg, Ted~W.
  Way, Dustin Werran, and Zhenbin Wu.
\newblock Fpga-accelerated machine learning inference as a service for particle
  physics computing.
\newblock {\em Computing and Software for Big Science}, 3(1):13, 2019.

\bibitem{CMS-DP-2024-121}
{CMS Collaboration}.
\newblock Model-independent real-time anomaly detection at the cms level-1
  calorimeter trigger with cicada.
\newblock Technical Report DP2024\_121, CMS Collaboration, 2024.
\newblock
  \url{https://twiki.cern.ch/twiki/bin/view/CMSPublic/L1TriggerDPGResults}.

\bibitem{autoencoder40LHC}
Ekaterina Govorkova, Ema Puljak, Thea Aarrestad, Thomas James, Vladimir Loncar,
  Maurizio Pierini, Adrian~Alan Pol, Nicol{\`o} Ghielmetti, Maksymilian
  Graczyk, Sioni Summers, Jennifer Ngadiuba, Thong~Q. Nguyen, Javier Duarte,
  and Zhenbin Wu.
\newblock Autoencoders on field-programmable gate arrays for real-time,
  unsupervised new physics detection at 40 mhz at the large hadron collider.
\newblock {\em Nature Machine Intelligence}, 4(2):154--161, 2022.

\bibitem{DeepLearning}
Ian Goodfellow, Yoshua Bengio, and Aaron Courville.
\newblock {\em Deep Learning}.
\newblock MIT Press, 2016.
\newblock \url{http://www.deeplearningbook.org}.

\bibitem{relu}
Kaiming He, Xiangyu Zhang, Shaoqing Ren, and Jian Sun.
\newblock Delving deep into rectifiers: Surpassing human-level performance on
  imagenet classification, 2015.

\bibitem{IEEE754_2019}
{IEEE}.
\newblock {IEEE Standard for Floating-Point Arithmetic}, 2019.

\bibitem{vivado}
Vivado design suite user guide, 2024.
\newblock Version 2024.1. Available:
  \url{https://www.xilinx.com/products/design-tools/vivado.html}.

\bibitem{xilinx_versal_2018}
Xilinx.
\newblock Introducing the versal architecture.
\newblock Technical report, Xilinx, 2018.

\bibitem{har70}
Aleksej Logacjov and Astrid Ustad.
\newblock {HAR70+}.
\newblock UCI Machine Learning Repository.
\newblock {DOI}: https://doi.org/10.24432/C5CW3D.

\bibitem{SWAMINATHAN2025906}
Tushar~Prasanna Swaminathan, Christopher Silver, Thangarajah Akilan, and
  Jitendra Kumar.
\newblock Benchmarking deep learning models on nvidia jetson nano for real-time
  systems: An empirical investigation.
\newblock {\em Procedia Computer Science}, 260:906--913, 2025.
\newblock Seventh International Conference on Recent Trends in Image Processing
  and Pattern Recognition (RTIP2R-2024).

\bibitem{technologies12060081}
Oumayma Jouini, Kaouthar Sethom, Abdallah Namoun, Nasser Aljohani,
  Meshari~Huwaytim Alanazi, and Mohammad~N. Alanazi.
\newblock A survey of machine learning in edge computing: Techniques,
  frameworks, applications, issues, and research directions.
\newblock {\em Technologies}, 12(6), 2024.

\bibitem{10.3389/fdata.2020.598927}
Yutaro Iiyama, Gianluca Cerminara, Abhijay Gupta, Jan Kieseler, Vladimir
  Loncar, Maurizio Pierini, Shah~Rukh Qasim, Marcel Rieger, Sioni Summers,
  Gerrit Van~Onsem, Kinga~Anna Wozniak, Jennifer Ngadiuba, Giuseppe
  Di~Guglielmo, Javier Duarte, Philip Harris, Dylan Rankin, Sergo Jindariani,
  Mia Liu, Kevin Pedro, Nhan Tran, Edward Kreinar, and Zhenbin Wu.
\newblock Distance-weighted graph neural networks on fpgas for real-time
  particle reconstruction in high energy physics.
\newblock {\em Frontiers in Big Data}, Volume 3 - 2020, 2021.

\bibitem{9043731}
Lei Deng, Guoqi Li, Song Han, Luping Shi, and Yuan Xie.
\newblock Model compression and hardware acceleration for neural networks: A
  comprehensive survey.
\newblock {\em Proceedings of the IEEE}, 108(4):485--532, 2020.

\bibitem{Qu_2020}
Huilin Qu and Loukas Gouskos.
\newblock Jet tagging via particle clouds.
\newblock {\em Physical Review D}, 101(5), March 2020.

\bibitem{Duarte_2018}
J.~Duarte, S.~Han, P.~Harris, S.~Jindariani, E.~Kreinar, B.~Kreis, J.~Ngadiuba,
  M.~Pierini, R.~Rivera, N.~Tran, and Z.~Wu.
\newblock Fast inference of deep neural networks in fpgas for particle physics.
\newblock {\em Journal of Instrumentation}, 13(07):P07027–P07027, July 2018.

\end{thebibliography}

%
%
%




\end{document}